\documentclass{ametsocV6.1}

\title{Probabilistic Spatial Interpolation of Sparse Data using Diffusion Models}

\authors{Valerie Tsao, \aff{a}\correspondingauthor{Valerie Tsao, valerie.tsao@duke.edu}
Nathaniel W. Chaney,\aff{a} 
Manolis Veveakis, \aff{a}
}

\affiliation{\aff{a}{Department of Civil and Environmental Engineering, Duke University, Durham, NC}}

\abstract{The large underlying assumption of climate models today relies on the basis of a "confident" initial condition, a reasonably plausible snapshot of the Earth for which all future predictions depend on. However, given the inherently chaotic nature of our system, this assumption is complicated by sensitive dependence, where small uncertainties in initial conditions can lead to exponentially diverging outcomes over time. This challenge is particularly salient at global spatial scales and over centennial timescales, where data gaps are not just common but expected. The source of uncertainty is two-fold: (1) sparse, noisy observations from satellites and ground stations, and (2) internal variability stemming from the simplifying approximations within the models themselves. 
\\
In practice, data assimilation methods are used to reconcile this missing information by conditioning model states on partial observations. Our work builds on this idea but operates at the extreme end of sparsity. We propose a conditional data imputation framework that reconstructs full temperature fields from as little as 1\% observational coverage. The method leverages a diffusion model guided by a prekriged mask, effectively inferring the full-state fields from minimal data points. We validate our framework over the Southern Great Plains, focusing on afternoon (12:00–6:00 PM) temperature fields during the summer months of 2018–2020. Across varying observational densities—from swath data to isolated in-situ sensors—our model achieves strong reconstruction accuracy, highlighting its potential to fill in critical data gaps in both historical reanalysis and real-time forecasting pipelines.
}

\begin{document}

%% Necessary!
\maketitle

%%%%%%%%%%%%%%%%%%%%%%%%%%%%%%%%%%%%%%%%%%%%%%%%%%%%%%%%%%%%%%%%%%%%%
% MAIN BODY OF PAPER
%%%%%%%%%%%%%%%%%%%%%%%%%%%%%%%%%%%%%%%%%%%%%%%%%%%%%%%%%%%%%%%%%%%%%

\section{Introduction} \label{intro}
In a world of data scarcity, researchers are frequently confronted with a long-standing question: how can we achieve comprehensive data coverage in regions where direct collection is limited or infeasible? This is not a new challenge, and is one that predates the digital era, rooted in a broader scientific aim to understand complex systems through incomplete information. Perhaps more importantly, the exercise of extrapolating from limited measurements enables us to make more informed decisions in areas regarding resource management, risk assessment, and policy frameworks.  In many cases, particularly within the climate and atmospheric sciences, data originates from field surveys that yield discrete, point-based observations. However, such point-based data are insufficient for developing the spatially continuous representations necessary to understand large-scale dynamics. Without these continuous datasets, our capacity to formulate nuanced and effective responses, ranging from disaster preparedness to ecological forecasting, is significantly constrained. Spatial interpolation offers a solution, enabling the estimation of values at unsampled locations by leveraging spatial correlations. Its utility spans a broad range of disciplines, including ecology, hydrology, geoscience, and even medical imaging.

Classically within the context of Earth System Models (ESMs) and numerical weather prediction (NWP), spatial interpolation techniques have been largely based on either deterministic methods, geostatistical methods, or some combination of the two. Inverse distance weighting (IDW) \citep{fotheringham1989spatial}, ordinary kriging \citep{Krige1951}, and regression kriging \citep{ODEH1995215} serve as respective canonical examples. While effective in many scenarios, these approaches often hinge on strong assumptions regarding stationarity, spatial autocorrelation, and smoothness, which can limit their flexibility and accuracy in heterogeneous environments.

A prevalent application of spatial interpolation methods and dealing with this issue of data sparsity is the concept of data assimilation. Developed in the 1960s, data assimilation is a concept that aims to provide an estimation of the state of a system by merging previous model estimates with present observations \citep{WangDA}. This process is performed sequentially so that model parameters are adjusted dynamically to minimize error and satisfy physical constraints typically prescribed by the type of problem. Assuming a stochastic formulation of both the model and observation, we can represent them as probability density functions $p(\cdot)$. Using a Bayesian framework, the goal of data assimilation is to derive a posterior distribution $p(x|y)$, as expressed by Bayes rule:
\begin{equation}\label{eq:bayes_da}
    p(x|y) = \frac{p(y|x) \cdot p(x)}{p(y)},
\end{equation}
where $p(x)$ refers to the information before assimilation described by the prior PDF, $p(y|x)$ refers to the likelihood of $y$ conditioned upon state $x$, and $p(y)$ refers to the probability of the state $y$ happening. In other words, this state estimation problem can be used to provide the model with an initial condition that can guide the future evolution of the solution in space and time.

Since its inception, data assimilation has evolved to encompass a suite of methodologies, including the Kalman filter (KF) \citep{KalmanWeather}, 4DVar \citep{4DVar}, ensemble Kalman smoother (EnKS) \citep{EnKS}, and hybrid forms which combine some form of the variational and ensemble Kalman filter (EnKF) method \citep{HybridEnKS}. These methods are now central to operational NWP and large-scale ocean modeling systems. Yet, traditional DA frameworks typically rely on simplifying approximations to achieve tractable expressions for analysis. As a result, they can struggle with bias correction, high-dimensional state spaces, varying sampling frequency, and integrating heterogeneous datasets with mismatched spatial and temporal resolutions.  The inherent complexity of tuning DA systems under conditions of multiscale, multivariate sparsity renders the “optimal” state estimation a perplexing target.

In parallel, recent advances for diffusion models, a small class of generative AI, has gained prominence for tasks in image and video generation \citep{nichol2022glidephotorealisticimagegeneration}, super-resolution \citep{rombach2022highresolutionimagesynthesislatent, 9887996}, and other computer vision applications \citep{chen2023diffusiondetdiffusionmodelobject, amit2022segdiffimagesegmentationdiffusion}. These models trace their conceptual roots to the thermodynamics-inspired Langevin dynamics proposed by \citet{Langevin}. Building on this foundation, \citet{SongErmon} introduced noise-conditioned score networks, which learn the gradient of the data distribution by training a model to match it directly. This line of work culminated in the formulation of denoising diffusion probabilistic models (DDPMs) by \citet{DDPM}, who advanced earlier efforts by discretizing the forward noising process into a Gaussian Markov chain and reparameterizing the reverse process to predict the added noise directly, yielding a simple and stable mean squared error objective during training. What makes diffusion models particularly promising in the context of spatial data reconstruction is their capacity to learn complex, high-dimensional data distributions without relying on restrictive parametric assumptions. Unlike classical DA, which often imposes physical or statistical constraints a priori, diffusion-based approaches learn these relationships implicitly from data. These realizations have sparked a growing body of research focused on combining diffusion models with data assimilation techniques for improved posterior estimation  \citep{huang2024diffdadiffusionmodelweatherscale, 10.2139/ssrn.4591077}.
% and transforming the training process to achieve greater accuracy. 
% expand more?? what is difference between DDPM and rest

Among the many generative tasks diffusion models can perform, inpainting, a technique for reconstructing missing or damaged portions of an image by leveraging contextual pixel information, has particular relevance for spatial data applications. The 2022 study by Lugmayr et al. demonstrated the potential of combining DDPMs with inpainting to reconstruct images through clever conditioning techniques \citep{RePaint}. This progress raises an intriguing question that underpins the present work: in a world of data sparsity, can diffusion models be harnessed for space-time interpolation in conditional simulations? 

To that end, this paper uses a combined framework of DDPM and kriging to produce reconstructed atmospheric fields for increasing levels of masked data and in-situ to swath ratios. In the following sections, 
a new framework is introduced, combining the generative abilities of DDPM and specific conditioning techniques to impute atmospheric fields at diverse levels of data sparsity. Following that, an in-depth comparison is attempted for this method against classical spatial interpolation methods such as inverse distance weighting (IDW) and conditional gaussian simulations (CGS). Finally, we investigate the potential to extend this framework into higher dimensionality, using 3D as a case study and critically discuss the results.

\section{Study Area \& Data}
Our study site lies at the intersection of the Central and Southern Great Plains region, encompassing the spatial extent between 36\textdegree 0' 23.4" to 37\textdegree 59' 1.0" N and 98\textdegree 58' 15.6" W to 96\textdegree 59' 33.0" W. This area is of particular interest because it experiences dramatic variability in climate conditions as a consequence of its complex topography and proximity to the Gulf of Mexico in the southeast. Broadly, the Great Plains exhibits a west-to-east gradient of increasing temperature and a north-to-south gradient of increasing precipitation. These gradients give rise to two distinct hydroclimatic regimes, setting the stage for a range of extreme weather events, including tropical cyclones, heatwaves, hailstorms, droughts, and blizzards. The region’s climatology is further shaped by the influence of the Rocky Mountains to the west, whose elevation and orientation contribute to a significant rain shadow effect, limiting the influx of Pacific moisture. Moreover, the confluence of cold, dry air masses from the north with warm, maritime air from the southeast creates highly dynamic atmospheric conditions, fostering the development of severe convective systems and frequent tornadic activity \citep{rosenberg1987climate}. In effect, both temperature and wind patterns in this region can shift dramatically over relatively short distances. According to projections from the Fifth National Climate Assessment (NCA5), intensifying aridity and temperature shifts driven by continued emissions are expected to exacerbate the frequency and severity of these extremes \citep{RN8287}. 

Given these strong local gradients in seasonal temperatures and the high interannual variability driven by opposing atmospheric influences, the region offers a rigorous testing ground for our proposed framework. By nature of its vastly diverse climate, the SGP is a location which holds much intrigue and scientific interest among the research community. To that end, it has been home to the largest atmospheric measurement site in the world, the Atmospheric Radiation Measurement (ARM) Facility, since the early 1990s \citep{TheARMSouthernGreatPlainsSGPSite}. It remains a data-rich environment, equipped with an extensive network of insitu and remote sensing instrumentation.

\begin{figure}[H]
    \centering
     \noindent\includegraphics [scale=0.37]{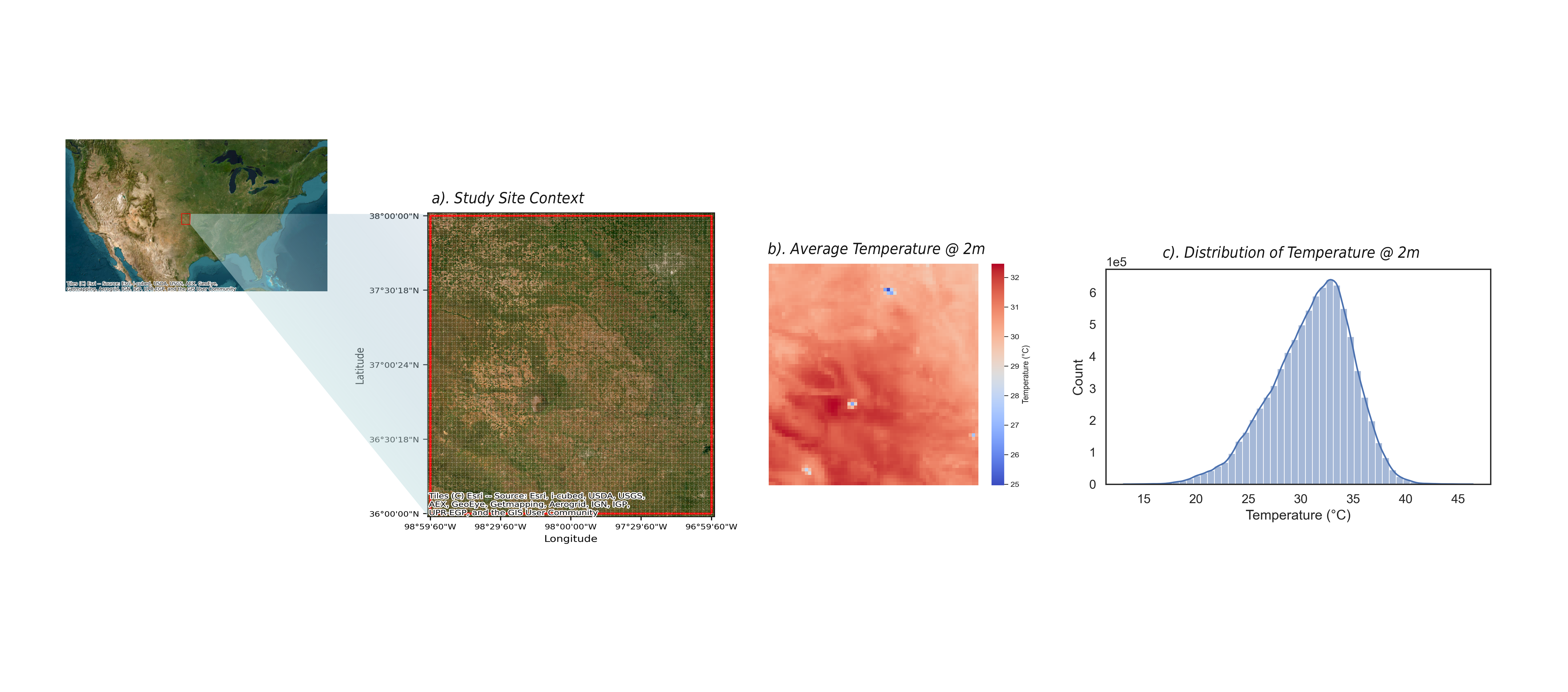}\\
     \caption{Study site in southeast Kansas extending into central Oklahoma, US. a) Geographic context of domain visualized using satellite imagery from ESRI. b) Temperature field averaged across entire study window. c) Distribution of all temperature values.}\label{fig: site}
\end{figure}

\subsection{Dataset} \label{data}
The dataset utilized in this study consists of historical U.S. temperature data at 2 meters above the surface, recorded hourly between 17:00 and 23:00 UTC during the summer months (June 1–August 31) from 2018 to 2020. These data were sourced from the National Oceanic and Atmospheric Administration (NOAA) High-Resolution Rapid Refresh (HRRR) model. The selected time window corresponds to 12:00–18:00 CST at our study site, capturing the critical midday to early evening period. While this study focuses on temperature fields, the proposed theoretical framework for kriging-smoothed conditioned diffusion (KrigSCD) can be easily generalized to other spatiotemporal datasets. Across the three summers, this totals to 6 hours/day $\times$ 92 days/summer $\times$ 3 summers = 1,656 hourly samples. From this set, 150 samples were randomly selected as our test set. 

\section{Methodology} \label{methodology}
\begin{figure}[H]
    \centering
     \noindent\includegraphics [scale=0.4]{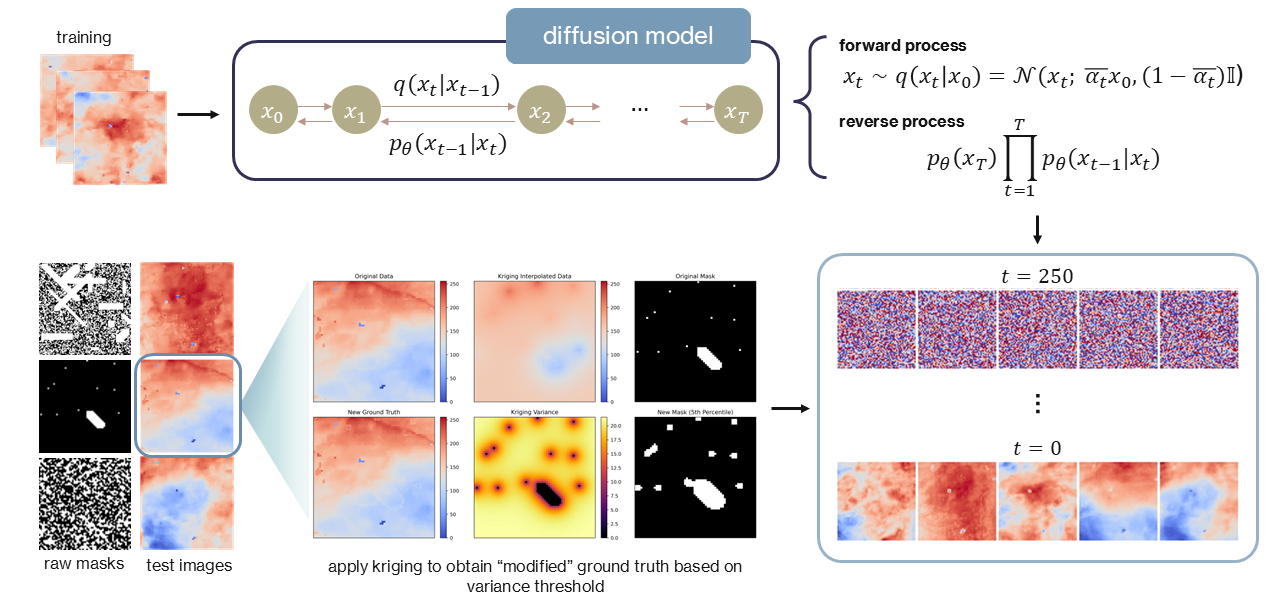}\\
     \caption{Schematic of proposed \textbf{Krig}ing-\textbf{S}moothed \textbf{C}onditional \textbf{D}iffusion (KrigSCD) framework.}\label{fig: framework}
\end{figure}

The following section outlines a comprehensive, multi-step methodology for reconstructing sparse 2D data fields from randomly simulated insitu and swath observations. Our approach is divided into three primary components: (1) a training process utilizing a diffusion model to learn the underlying distribution of input images; (2) a mask generation process that simulates realistic observation patterns, where individual pixels represent in-situ data and randomized trajectories of varying lengths and directions correspond to satellite swath observations; and (3) a guiding process that refines the model’s focus with something we refer to as a kriged smoother, directing it to converge around regions of observational coverage. This process is depicted in Figure \ref{fig: framework}. Finally, we assess the performance of the model on previously unseen, out-of-sample images and quantify its performance against traditional methods.

\subsection{Problem Formulation} \label{problem}
The problem of deriving a plausible structure of the atmosphere with minimal data can be formally defined as follows:

Let images be defined on a square grid of size \(N\times N\), and let \(\mathbf{x}\in\mathbb{R}^{N^2}\) denote the vectorized true image.  A binary mask operator \(H\in\{0,1\}^{M\times N^2}\) selects the \(M\) observed pixels so that the data vector is  
\[
\mathbf{y} \;=\; H\,\mathbf{x} \;+\;\boldsymbol\eta,\quad \boldsymbol\eta\sim\mathcal{N}(\mathbf{0},\mathbf{R}),
\]  
where \(\mathbf{R}\in\mathbb{R}^{M\times M}\) is the observation–error covariance.  Our goal is to reconstruct the missing entries of \(\mathbf{x}\) (i.e.\ the \(N^2 - M\) unobserved pixels) by enforcing consistency with \(\mathbf{y}\) and incorporating spatial‐correlation priors.  Equivalently, we seek the analysis state  
\[
\mathbf{x}^a \;=\;
\arg\min_{\mathbf{x}\in\mathbb{R}^{N^2}}
\;\frac12(\mathbf{x}-\mathbf{x}^b)^\top\mathbf{B}^{-1}(\mathbf{x}-\mathbf{x}^b)
\;+\;
\frac12(H\mathbf{x}-\mathbf{y})^\top\mathbf{R}^{-1}(H\mathbf{x}-\mathbf{y}),
\]
where \(\mathbf{x}^b\in\mathbb{R}^{N^2}\) is a background (prior) image and \(\mathbf{B}\in\mathbb{R}^{N^2\times N^2}\) its covariance.  Classical interpolation schemes arise as particular choices of \(\mathbf{B}\) (e.g.\ ordinary kriging from a variogram‐based \(\mathbf{B}\), inverse‐distance weighting from a diagonal \(\mathbf{B}\)), thus embedding the task of mask‐based image imputation within a variational data‐assimilation framework. We compare these classical methods with a relatively new but popular diffusion framework.

\subsection{Diffusion Models} \label{diffusion_model}
Diffusion models are fundamentally generative models that work by converting a complex distribution (describing a set of input images) into a simpler one, analogous to processes used in GANs \citep{goodfellow2014generativeadversarialnetworks}, VAEs \citep{kingma2015autoencodingvariationalbayes}, and normalizing flows \citep{rezende2016variationalinferencenormalizingflows}. This is done through a 2-step approach--a forward process, where noise is added iteratively at each timestep until we reach a final predefined final timestep $T$; and a reverse process, where a neural network learns the mean $\mu_\theta$ and variance $\Sigma_\theta$ of an approximated conditional probability distribution we'll call $p_\theta (x_{t-1} | x_t)$ that describes the "denoising" process. This allows us to go from pure noise to a sensible image. Our explanation here is rather high-level and will be further refined more mathematically in the ensuing paragraphs, but the main goal here is to provide some intuition for how these models work and interact. An alternate perspective we can take is to view these diffusion models in the context of an image space. If we imagine an image space being the set of all possible images that can exist, where every pixel in a $n \times n$ image is defined by a value between $0-255$, it should make intuitive sense that most of this nebulous space is nonsensical, with few clusters of things we define as "good images". Therefore, during inference, our diffusion model is essentially pushing a single sample that lies in a cluster of temperature field images outside of the cluster boundaries during the forward process; then, attempting to learn the parameters that will allow it to trace back to the underlying distribution that describes temperature field images in the reverse process. 

To understand how we can train diffusion models to denoise, we provide a formulation of Gaussian diffusion models as given by the seminal paper by \citet{DDPM}. We are given an initial $d-$dimensional image drawn from some probability distribution $q(x)$: $x_0 \in \mathbb{R}^{n_1\times n_2 \times \cdots \times n_d} \sim q(x)$, represented by a $d-$dimensional tensor with dimensions $n_1\times n_2\times \cdots \times n_d$. Here, $q$ is some arbitrary Markovian noising process -- it gradually adds noise to the data to produce ``noised'' samples $x_1,x_2,\dots, x_T$. This is what we referred to as the \textit{forward diffusion process} earlier. That is, for each timestep $1 \leq t \leq T$, 
\[q(x_t | x_{t-1}) = \mathcal{N}\Big(x_t, \sqrt{1-\beta_t} x_{t-1}, \beta_t \mathbf{I}\Big).\]
Here, $\beta_t$ is some variance schedule (often chosen to be linear or sinusoidal in $t$). Ideally we would like to find $q(x_T)$, but this proves to be difficult for large $T$, as we find that
\begin{equation}\label{eq:fwd}
    q(x_{1:T}|x_0)=\prod_{t=1}^T q(x_t|x_{t-1}).
\end{equation}

By reparameterizing using $\alpha_t=1-\beta_t$ and $\bar{\alpha}_t=\prod_{s=1}^t \alpha_s$ with the noise $\epsilon_t\sim \mathcal{N}(0, \mathbf{I})$, $x_t$ follows a closed form
\begin{equation}\label{eq:reparam}
    x_t=\sqrt{\bar{\alpha}_t}x_0+\sqrt{1-\bar{\alpha}_t} \epsilon_0,
\end{equation}
and so to produce a sample at any timestep $t$ it suffices to sample from $x_t\sim q(x_t|x_0)=\mathcal{N}(x_t;\sqrt{\bar{\alpha}_t}x_0,(1-\bar{\alpha}_t)\mathbf{I})$. As $\beta_t$ is predetermined, we can precompute our $\alpha$ coefficients for all timesteps.

When taking $T\to\infty$ (and assuming some regularity conditions on $\beta_t$), $x_T$ approaches an isotropic Gaussian distribution. This means that if we wish to sample from $q(x_0)$, it suffices to sample from $q(x_T)$ and then sample reverse steps $q(x_{t-1}|x_t)$ until we reach $x_0$, which is what we referred to as the \textit{backward diffusion process} from before. However, in practice $q(x_{t-1}|x_t)$ is intractable, as finding a closed form would require information about $q(x_0)$, which we do not know. Thus, we must approximate it with a parameterized model $p_\theta$, which in our case is a neural network. Knowing that $q(x_{t-1}|x_t)$ is Gaussian, we can choose the mean $\mu_\theta$ and covariance $\Sigma_\theta$ matrices as our learned parameters in $p_\theta$:
\begin{equation}\label{eq:bwd_nn}
p_\theta(x_{t-1}|x_t)=\mathcal{N}\Big(x_{t-1};\mu_\theta(x_t,t),\Sigma_\theta(x_t,t)\Big).
\end{equation}
$\Sigma_\theta(x_t, t)$ is learned is through an interpolation between $\beta_t$ and $\Tilde{\beta}_t=\frac{1-\bar{\alpha}_{t-1}}{1-\bar{\alpha}_t}\cdot \beta_t$ by predicting a mixed model vector $v$:
\begin{equation}\label{eq:learn_sigma}
    \Sigma_\theta(x_t,t) = \exp\left(v\log \beta_t + (1-v)\log \Tilde{\beta}_t\right).
\end{equation}
Now, analogously to the forward diffusion process found in Eq. \ref{eq:fwd}, the backward diffusion process is given by
\begin{equation}\label{eq:bwd}
    p_\theta(x_{0:T})=p_\theta(x_T)\prod_{t=1}^T p_\theta(x_{t-1}|x_t).
\end{equation}
To train the model such that $p(x_0)$ learns the data distribution $q(x_0)$, we follow the techniques found in variational autoencoders (which diffusion models happen to be equivalent to). We aim to minimize the variational lower bound $\mathcal{L}_{\text{vlb}}$ given by 
\begin{equation}\label{eq:vlb}
    \mathcal{L}_{\text{vlb}} = \mathcal{L}_0 + \underbrace{\mathcal{L}_1 + \cdots + \mathcal{L}_{T-1}}_{\mathcal{L}_t} + \mathcal{L}_T.
\end{equation}
Here,
\begin{align*}
    \mathcal{L}_0 &\coloneqq -\log p_\theta(x_0|x_1), \\
    \mathcal{L}_t &\coloneqq D_{\text{KL}}(q(x_{t-1}|x_t, x_0) \mid\mid p_\theta(x_T)), \\
    \mathcal{L}_{T} &\coloneqq D_{\text{KL}}(q(x_T|x_0) \mid \mid p(x_T)),
\end{align*}
where $D_{\text{KL}}(p \mid\mid q) $ is the KL-divergence between two probability distrubitions $p$ and $q$. Intuituively, $\mathcal{L}_0$ follows the reconstruction term found in the ELBO of a variational autoencoder, and in practice is learned using a separate decoder. $\mathcal{L}_t$ formulates the difference between the approximated denoising steps $p_\theta(x_{t-1}|x_t)$ and the desired ones $q(x_{t-1}|x_t, x_0)$. $\mathcal{L}_T$ shows how close $x_T$ is to the standard Gaussian.

While this objective is well-justified, it was found in \citep{DDPM} that a simpler objective performs better in practice. In particular, they do not directly parameterize $\mu_\theta(x_t, t)$ as a neural network, but rather train a model $\epsilon_\theta(x_t,t)$ to predict the noise $\epsilon$ from Equation \ref{eq:reparam}. This transforms our loss into 
\begin{equation}\label{eq:simp_loss}
    \mathcal{L}_{\text{simple}} = \mathbb{E}_{t \sim [1,T], x_0\sim q(x_0), \epsilon \sim \mathcal{N}(0, 1)}\Big[\|\epsilon - \epsilon_\theta(x_t, t)\|^2\Big].
\end{equation}
During sampling, we can use substitution to derive $\mu_\theta(x_t, t)$ from $\epsilon_\theta(x_t, t)$:
\[\mu_\theta(x_t, t)=\frac{1}{\sqrt{\alpha_t}}\left(x_t - \frac{1-\alpha_t}{\sqrt{1-\bar{\alpha}_t}}\epsilon_\theta(x_t, t)\right).\]
Finally, when we are learning the covariance $\Sigma_\theta$, we must also modify the loss function to incorporate $\Sigma_\theta$, in which case we would have 
\[\mathcal{L}_{\text{final}} = \mathcal{L}_{\text{simple}} + \lambda \mathcal{L^\prime}_{\text{vlb}},\]
where $\mathcal{L^\prime}_{\text{vlb}}$ is a modified version that only learns $\Sigma_\theta$, and $\lambda=0.001$ is a small tuning parameter. 

\subsection{Mask Generation} \label{mask_generation}
In order to carry out the task of spatiotemporal data imputation in the context of image inpainting, we must provide a binary mask of known and unknown data values for which the diffusion model can use as guidance for the denoising process. These masks are generated in such a way as to mimic observations that would be taken in a real-word setting, namely insitu and satellite swaths. Insitu observations are represented as randomly distributed isolated pixels, while satellite swaths are modeled as linear segments with fixed width, variable length, and randomized direction. This allows us to capture a wide distribution of customizable satellite configurations. A desired target percentage of known data for the overall image is set beforehand along with the the ratio between insitu and swath observations, taken as a value between 0$-$1. This ensures flexible and realistic mask generation tailored to a range of observational scenarios, useful for training or evaluating models under controlled yet representative sampling conditions.

\subsection{Mask Conditioning for Interpolation} \label{mask_conditioning}
For the purposes of image inpainting, \citep{RePaint} used the following idea: to predict missing pixels in an image, one can use the mask region as a condition. Since at each reverse step $x_{t}\to x_{t-1}$ only depends on $x_t$, we can keep the ``correct'' properties of $q(x_t)$. In other words, rather than generating $q(x_{t-1})$ uniformly, we decompose the sample into \textit{known} and \textit{unknown} regions, each having a different distribution.

In this section, we shall denote the grouth truth image as $x$, the unknown pixels as $m\odot x$ and the known pixels as $(1-m)\odot x$, where $\odot$ refers to the Hadamard product. Then 
\begin{align*}
    x_{t-1}^{\text{known}} &\sim \mathcal{N}\Big(\sqrt{\bar{\alpha}_t}x_0, (1-\bar{\alpha}_t)\mathbf{I}\Big), \\
    x_{t-1}^{\text{unknown}} &\sim \mathcal{N}(\mu_\theta(x_t, t), \Sigma_\theta(x_t, t)), \\
    x_{t-1} &= m \odot x_{t-1}^{\text{known}} + (1-m) \odot x_{t-1}^{\text{unknown}}.
\end{align*}
However, the authors showed that na{\"i}vely applying this yielded images that were discombobulated; the sampling of the known pixels is performed without considering the generated parts of the image, which obstructs any potential synergy between the two. To allow more time for the conditional known pixels and the generated unknown pixels to harmonize, a resampling method is enacted by diffusing the output $x_{t-1}$ back into $x_t$ using the same forward process $x_t \sim \mathcal{N}(x_t;\sqrt{\bar{\alpha}_t}x_0,(1-\bar{\alpha}_t)\mathbf{I})$. This is done $r$ times before continuing to the next time step. While this process introduces noise to the known regions, some information from the generated $x_{t-1}^{\text{unknown}}$ is preserved in $x_{t}^{\text{unknown}}$, leading to a better $x_{t}^{\text{unknown}}$ overall that is more aligned with the information contained in $x_{t}^{\text{known}}$. This resampling is performed every $j$ timesteps.
\subsection{KrigSCD} \label{krigscd}
When applying the conditioning discussed above to our dataset, we found that the model still struggled to effectively utilize insitu point observations. While it's true that individual known pixels would align with "ground truth" sample values post-inference, the model was clearly not propagating these types of observations to adjacent pixels nor fully taking advantage of the known information. To combat this, we introduced a smoothing process that utilizes ordinary kriging realizations to revise the original binary mask based on a set threshold, such that any known insitu observations will not be read in as isolated points of known data but rather tightly concentrated regions of known. 
\begin{figure}[H]
\centering
 \noindent\includegraphics [width=19pc,angle=0]{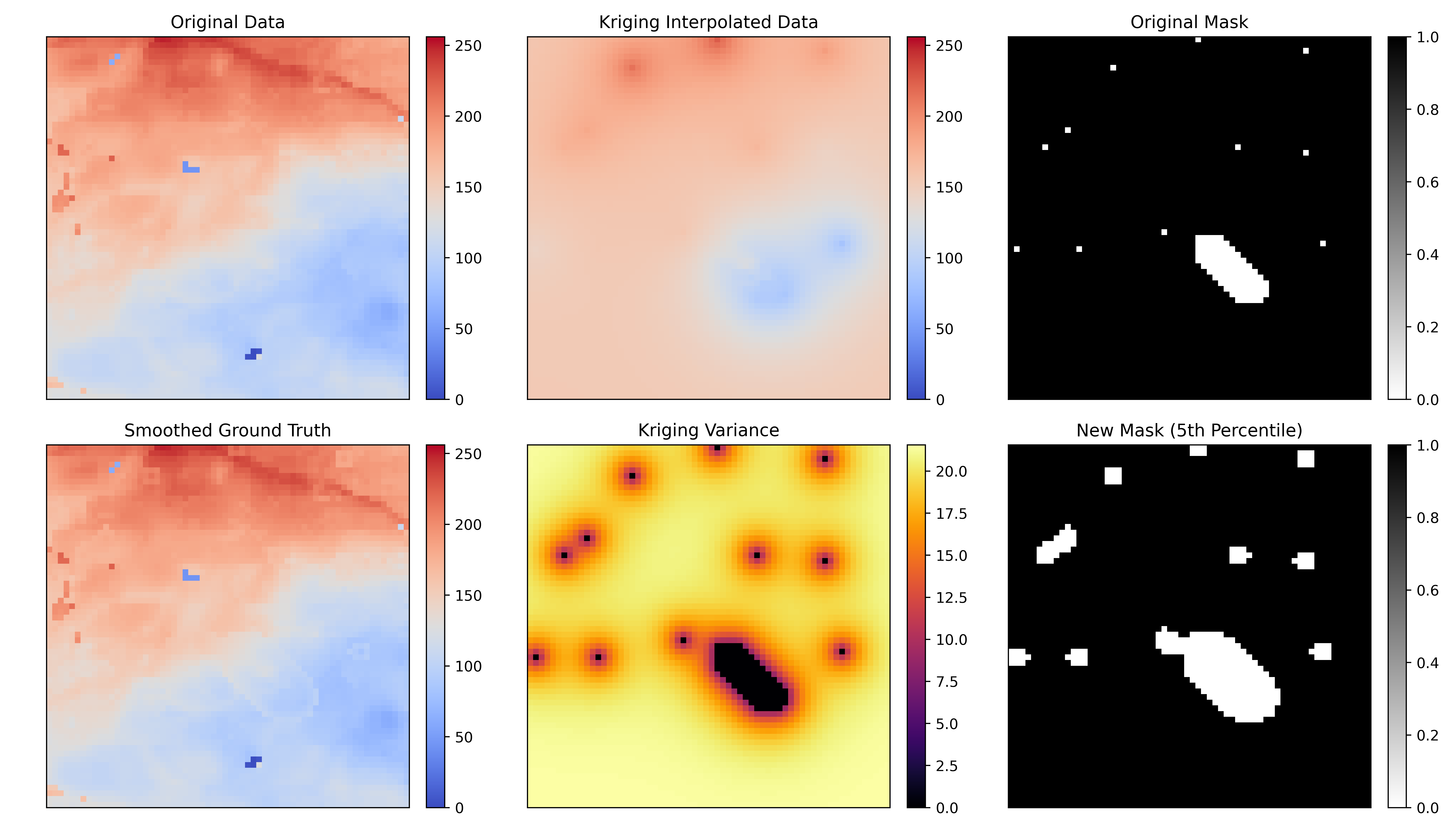}\\
 \caption{A before/after look at the proposed smoothing process. Starting at the top left and going clockwise, we have the original ground truth, the interpolated image using ordinary kriging, the original mask used to generate the interpolated image, the updated mask fed into the diffusion model based on kriging variance percentiles, the kriging variance of the interpolated image, and the updated ground truth fed into the diffusion model.}\label{fig: krigscd}
\end{figure}

In the proposed KrigSCD (\textbf{Krig}ing-\textbf{S}moothed \textbf{C}onditional Diffusion) approach highlighted in Figure \ref{fig: krigscd}, pixels in the masked (unknown) regions are replaced by their kriged estimates $z^*$ only if the corresponding kriging standard deviation $\sigma$ is below a predefined threshold (which we chose as the 5th percentile among all unknown pixels). The mask is updated accordingly, producing refined ground truth and mask pairs that are then used as input to a diffusion model for inpainting. This pre-processing step leverages spatial correlations captured via the semivariogram and ordinary kriging to reduce uncertainty in the inpainting task, leading to improved reconstruction quality. However, when computing the errors, we make sure to compare the results to that of the original, pre-smoothed ground truth. Below, we present the algorithm for data imputation using the unconditional DDPM with a mask condition during the reverse process. 

\begin{algorithm}[H]
  \caption{Ordinary Kriging}
  \label{alg:ordinary_kriging}
  \begin{algorithmic}[1]
    \REQUIRE Known samples $\{(\mathbf{x}_i,\,z_i)\}_{i=1}^n$,  
             target locations $\{\mathbf{x}^*_j\}_{j=1}^m$,  
             variogram model $\gamma(h)$
    \ENSURE Estimates $\{\hat z^*_j\}_{j=1}^m$
    \FOR{$j \gets 1\ \mathbf{to}\ m$}
      \STATE 
        $h_{ij} \gets \|\mathbf{x}_i - \mathbf{x}^*_j\|,\ i=1,\dots,n$
      \STATE 
        $\boldsymbol{\gamma}_j \gets [\,\gamma(h_{1j}),\dots,\gamma(h_{nj}),\,1\,]^\top$
      \STATE 
        $\Gamma_{ik} \gets \gamma\bigl(\|\mathbf{x}_i-\mathbf{x}_k\|\bigr)$ 
        for $i,k=1,\dots,n$
      \STATE
          $\Gamma' \;=\;
          \begin{bmatrix}
            \Gamma & \mathbf{1}\\
            \mathbf{1}^\top & 0
          \end{bmatrix},\quad
          \boldsymbol{\gamma}'_j \;=\;
          \begin{bmatrix}
            \boldsymbol{\gamma}_j(1\!:\!n)\\
            1
          \end{bmatrix}$
          
      \STATE 
        $\Gamma'\,\boldsymbol{\lambda}'_j = \boldsymbol{\gamma}'_j$
      \STATE
        $\lambda_{ij} = \boldsymbol{\lambda}'_j(i),\ i=1,\dots,n$
      \STATE 
        $\displaystyle \hat z^*_j = \sum_{i=1}^n \lambda_{ij}\,z_i$
    \ENDFOR

    \FOR{$j \gets 1\ \mathbf{to}\ m$}
        \IF{$\operatorname{Var}(\hat{z}_j^*) \leq P_5\Big(\{\operatorname{Var}(\hat{z}_j^*)\}_{j=1}^m \Big)$}
            \STATE $(\mathbf{x}_j, z_j) = (\mathbf{x}_j^*, \hat{z}_j^*)$
            \ENDIF
    \ENDFOR
    \RETURN $\{\hat z^*_j\}_{j=1}^m$
  \end{algorithmic}
\end{algorithm}

\begin{algorithm}[H]
  \caption{Mask Conditioning}
\label{alg:inpainting}
\begin{algorithmic}[1]
    
    \STATE $x_T \sim \mathcal{N}(0, I)$
    \FOR{$t = T$ \TO $1$}
        \FOR{$u = 1$ \TO $U$}
            \IF{$t > 1$}
                \STATE $\epsilon_t \sim \mathcal{N}(0, I)$
            \ELSE
                \STATE $\epsilon_t = 0$
            \ENDIF
            \STATE $x_{t-1}^\text{known} \gets \sqrt{\bar{\alpha}_t} \, x_t + \left(\sqrt{1 - \bar{\alpha}_t}\right)\,\epsilon_t$
            \STATE $z \sim \mathcal{N}(0,I)$
            %\STATE \textcolor{blue}{// Here, we combine known parts of the input (masked or unmasked) appropriately}
            \STATE $x_{t-1}^\text{unknown} \gets \frac{1}{\sqrt{\alpha_t}}\Bigl(x_t - \beta_t \/\sqrt{1-\bar{\alpha}_t}\,\epsilon_\theta(x_t,t)\Bigr) + \sigma_t z$
            \STATE $x_{t-1} = m \otimes x_{t-1}^\text{known} + (1 - m) \otimes x_{t-1}^\text{unknown}$
            \IF{$u < U$ \AND $t > 1$}
                \STATE $x_t \sim \mathcal{N}\bigl(\sqrt{1-\beta_t}\,x_{t-1}, \beta_t I\bigr)$
            \ENDIF
        \ENDFOR
    \ENDFOR
    \RETURN $x_0$
\end{algorithmic}
\end{algorithm}

\subsection{Metrics} \label{metrics}
To evaluate the performance of our framework, we use a number of metrics which are described in more detail below. 
\subsubsection{Root Mean Squared Error} \label{rmse}
A common error metric is the \textit{root mean squared error} (RMSE):
\begin{equation}\label{eq:RMSE}
    \mathrm{RMSE} = \sqrt{\frac{1}{N}\sum_{i=1}^N (\hat{x}_i - x_i)^2},
\end{equation}
where $N$ is the number of non‑missing pixels. RMSE quantifies the standard deviation of the reconstruction errors, penalizing larger deviations quadratically and thus emphasizing outliers in the reconstruction.

\subsubsection{Mean Absolute Error} \label{mae}
The \textit{mean absolute error} (MAE) is defined as
\begin{equation}\label{eq:MAE}
    \mathrm{MAE} = \frac{1}{N} \sum_{i=1}^N \bigl|\hat{x}_i - x_i\bigr|,
\end{equation}
where $N$ is the number of pixels, $x_i$ is the ground truth intensity, and $\hat{x}_i$ is the reconstructed intensity. MAE measures the average absolute deviation in the original intensity units, offering an interpretable metric that is more robust to outliers than RMSE.

\subsubsection{Mean Relative Error} \label{mre}

For greyscale images with pixel intensities ranging from 0 to 255, the mean relative error (MRE) quantifies the average discrepancy between the ground truth and reconstructed images on a normalized scale. Let \(I_{\text{true}}(i)\) and \(I_{\text{recon}}(i)\) represent the intensity of the \(i\)th pixel in the ground truth and reconstruction, respectively, and let \(N\) be the total number of pixels. Then the MRE is given by
\begin{equation}\label{eq:MRE}
\text{MRE} = \frac{1}{N} \sum_{i=1}^{N} \frac{ I_{\text{true}}(i) - I_{\text{recon}}(i)}{255}.
\end{equation}
This metric yields a value between $-1$ and $1$, where values closer to zero indicate a reconstruction that is closer to the ground truth in terms of pixel intensities.

\subsubsection{Learned Perceptual Image Patch Similarity} \label{lpips}
The Learned Perceptual Image Patch Similarity (LPIPS) metric measures the perceptual similarity between two images by comparing deep features extracted from a pretrained convolutional neural network \citep{zhang2018unreasonableeffectivenessdeepfeatures}. Unlike simple pixel-wise comparisons, LPIPS leverages features from networks (which in our case is AlexNet), which capture higher-level details that are closer to human visual perception.

Let \( x \) and \( y \) be two images (for example, a ground truth and its interpolated version). For each layer \( l \) in the network, we extract feature maps
\[
\phi_l(x) \in \mathbb{R}^{H_l \times W_l \times C_l},
\]
where \( H_l \), \( W_l \), and \( C_l \) are the height, width, and number of channels, respectively. These feature maps are normalized channel-wise:
\[
\hat{\phi}_l(x) = \frac{\phi_l(x)}{\|\phi_l(x)\|_2},
\]
and similarly for \( y \).

The LPIPS distance is then defined as:
\begin{equation}\label{eq:lpips}
\text{LPIPS}(x,y) = \sum_{l=1}^{L} \frac{1}{H_l W_l} \sum_{h=1}^{H_l} \sum_{w=1}^{W_l} \left\| w_l \odot \left( \hat{\phi}_l(x)_{hw} - \hat{\phi}_l(y)_{hw} \right) \right\|_2^2,
\end{equation}
where:
\begin{itemize}
  \item \( \hat{\phi}_l(x)_{hw} \) and \( \hat{\phi}_l(y)_{hw} \) are the normalized feature vectors at location \((h, w)\) in layer \( l \),
  \item \( w_l \) is a learned weight vector for the \( l \)th layer,
  \item \( \odot \) denotes element-wise multiplication.
\end{itemize}

For greyscale images, the single channel is typically replicated across three channels to ensure compatibility with the network’s expected RGB input. Both images are normalized as required by the network before feature extraction.

This formulation of LPIPS provides a robust way to quantify perceptual differences, focusing on high-level structural and textural details rather than mere pixel-level variations.

\subsubsection{Lacunarity} \label{lacunarity}
Lacunarity measures the texture heterogeneity of an image by quantifying the distribution of gaps or voids. Given the true lacunarity values computed on the ground truth image, denoted as \(\Lambda_{\text{true}}(s)\) for different scales \(s\), and the corresponding lacunarity values from the reconstructed image, \(\Lambda_{\text{recon}}(s)\), the lacunarity norm error is defined as
\begin{equation}\label{eq:lac}
E_{\Lambda} = \sqrt{\sum_{s=1}^{S} \left(\Lambda_{\text{true}}(s) - \Lambda_{\text{recon}}(s)\right)^2}.
\end{equation}
This error provides a single measure of the deviation in texture patterns across all scales, with lower values indicating closer agreement between the reconstruction and the ground truth.

\subsubsection{Kernel Inception Distance} \label{kid}
The \textit{Kernel Inception Distance} (KID), assesses the distance between real and generated image distributions using the squared Maximum Mean Discrepancy (MMD) on features extracted by a pretrained Inception network \citep{bińkowski2021demystifyingmmdgans}. Let $\{\mathbf{f}_i\}_{i=1}^n$ and $\{\mathbf{g}_j\}_{j=1}^m$ be the $d$‑dimensional activation vectors of real and generated images, respectively, at a chosen layer. The unbiased estimator of squared MMD is
\begin{equation}\label{eq:KID}
    \mathrm{KID}^2 = \frac{1}{n(n-1)}\sum_{i\neq i'} k(\mathbf{f}_i,\mathbf{f}_{i'}) 
    + \frac{1}{m(m-1)}\sum_{j\neq j'} k(\mathbf{g}_j,\mathbf{g}_{j'}) 
    - \frac{2}{nm}\sum_{i,j} k(\mathbf{f}_i,\mathbf{g}_j),
\end{equation}
where the kernel is taken as
\[
k(\mathbf{u},\mathbf{v}) = \biggl(\frac{1}{d}\,\mathbf{u}^\top \mathbf{v} + 1\biggr)^3.
\]
KID yields nonnegative values, with smaller scores indicating closer agreement between the generated and real feature distributions. This is mainly used in measuring how well a trained diffusion model is at generating images that match the training (or input) distribution.

\subsection{Model Setup} \label{model_setup}
To train, we employ the guided diffusion model developed by OpenAI, which utilizes the Adam optimizer and a learning rate of $\alpha=\texttt{3e-4}$. 
The architecture of the model follows \citet{guided-diffusion}, which leverages a UNet backbone incorporating residual blocks and a combination of downsampling and upsampling convolutions. The only main changes made include using attention at 64$\times$64 on top of the 32$\times$32,  16$\times$16,  and 8$\times$8 resolutions from the original model. We make use of the variance scheduler (i.e., setting $\texttt{learn\_sigma=True}$), as prior experiments with fixed variance consistently produced disjointed and visually inconsistent outputs that failed to align closely with ground truth distributions. On one GPU (NVIDIA RTX A5000) and batch size 1, our model was trained for 100k epochs and 250 diffusion steps, which took roughly 10 hours on images of size 64x64. 

For the mask-conditioned diffusion step, we chose $r=10$ resampling steps and $j=10$ resampling frequency, with $t_T=150$ timestep re-spacing. Due to the probabilistic nature of our algorithm, we conducted a study to determine the adequate number of ensemble members to further refine our results, shown below in Figure \ref{fig:ensemble_number}. 
\begin{figure}[h]
    \centering
    \includegraphics[scale=0.27]{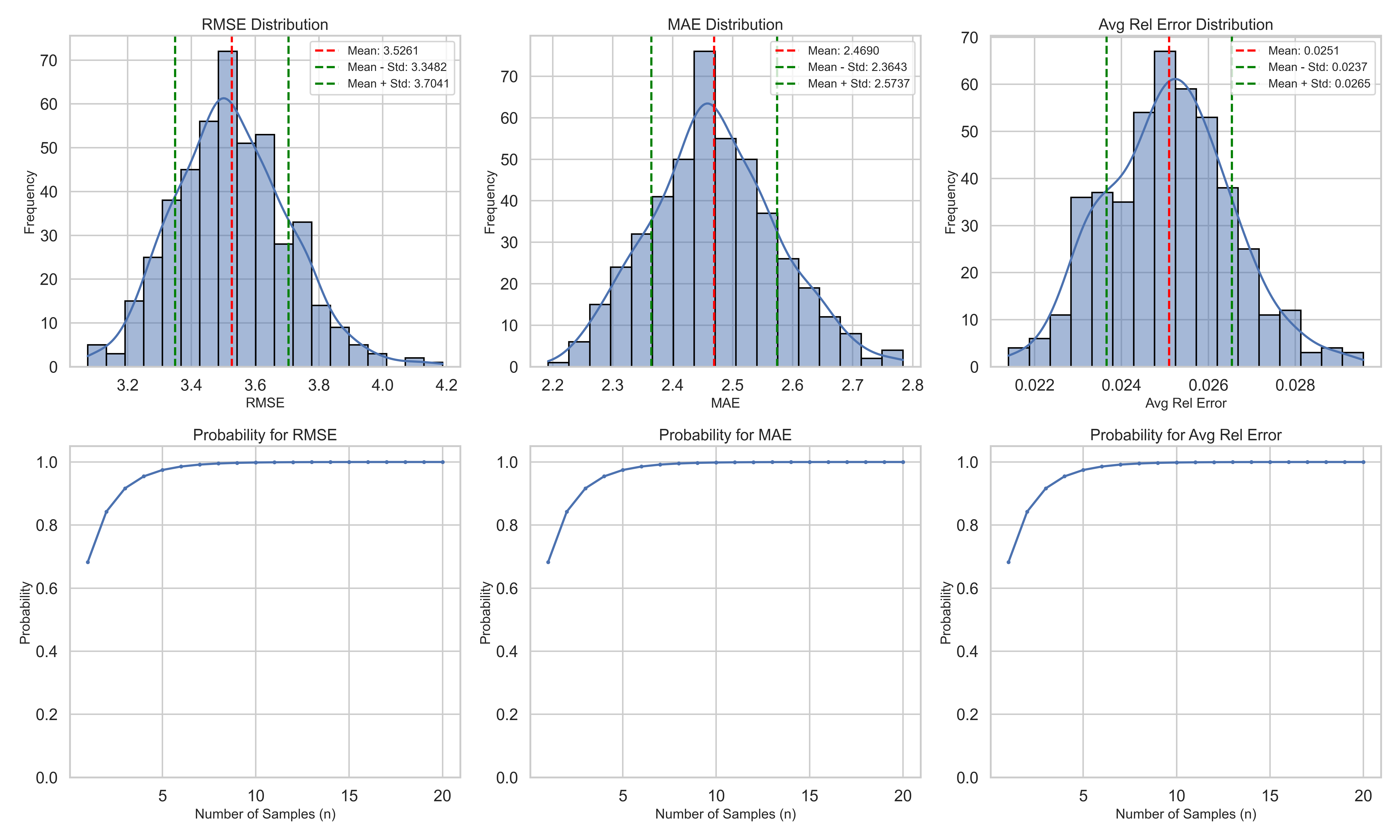}
    \caption{A graph of three error metrics for diffusion spatial interpolations: RMSE, MAE, and MRE. On the histogram we denote the mean and first standard deviation of the metrics by the red and green dotted lines respectively, and below each graph the probability that $n$ samples will lie within one standard deviation of the mean.} 
    \label{fig:ensemble_number}
\end{figure}

With $n=500$ samples, we satisfy the requirements for the Central Limit Theorem and can assume that our distribution is approximately Gaussian. Thus using a classical result from statistics, where if $\bar{x}$ denotes our sample mean, we know that $\bar{x}\sim \mathcal{N}(\mu, \sigma/\sqrt{n})$, and so the probability that $\bar{x}$ lies within one standard deviation of the true mean is given by
\[\mathbb{P}(|\bar{x}-\mu|<\sigma) = 2\Phi(\sqrt{n}) - 1,\]
where $\Phi(\cdot)$ is the CDF of the standard Gaussian distribution. Using this formula, we deduce that a size of $n=10$ ensemble members is sufficient to obtain a reconstruction that is close enough (i.e. within one standard deviation) to the true mean error of each distribution of reconstructions. In order to remain consistent to give the best possible comparison, we also chose $n=10$ ensemble members for kriging with conditional Gaussian simulations.

\section{Results} \label{results}
\subsection{Unconditional Diffusion Generations} \label{uncond)diff}
\begin{figure}[H]
    \centering
    \includegraphics[scale=0.2]{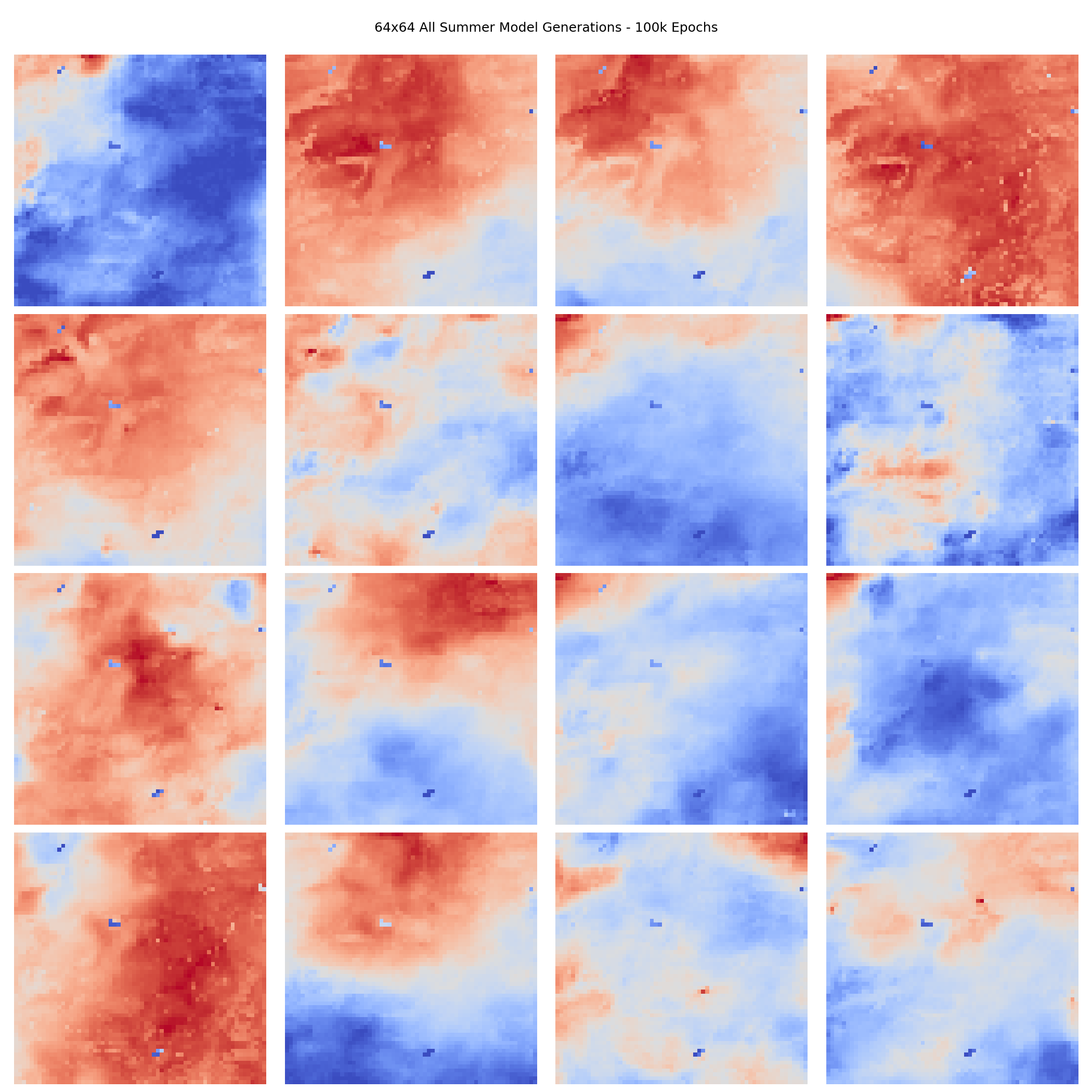}
    \caption{16 sample generations from the unconditioned diffusion model trained on our dataset. These are not conditioned on any mask coverage and are meant to act as a quality check of the diffusion component before proceeding with the conditional sampling process. As shown, the model is correctly learning where the cities are, which are denoted by the dark blue pixels and appear at the exact same locations in each generation.}
    \label{fig:uncond_generations}
\end{figure}

To preface the output of KrigSCD, we first show the generations from the base diffusion before any mask conditioning or resampling techniques have been applied. Figure \ref{fig:uncond_generations} captures a random sample of these unconditional generations after training the model on 100k epochs. The significance in checking this before proceeding with the proposed framework allows us to judge the quality of the model to see if it produces images that could have feasibly been drawn from the ground truth image distribution. If these generations did not remotely match images that we see within our training, then we would know that the diffusion model is not sufficiently learning the features that define our samples. 

In order to quantify the quality of these generations against our ground truth images, we calculated the KID metric defined in Section \ref{metrics}\ref{kid}. Doing this with models trained at 100k, 150k, and 200k epochs yielded respective KID scores of 0.0798, 0.0640,  0.0418. Recall that values closer to 0 indicate greater similarity between the features of the generated and actual image distributions. Given that the number achieved for our 100k epoch model is already considered sufficiently low, we stick to the 100k model for simplicity but reasonably infer that using a model trained with higher epochs would result in even greater advantages in reconstruction accuracy. Note that to have an interpretable value for the KID score, we produce roughly the same amount of generations as samples in our training. 

\subsection{KrigSCD Reconstructions} \label{krigscd_reconstructions}
\begin{figure}[H]
\centering
  \includegraphics[scale=0.4]{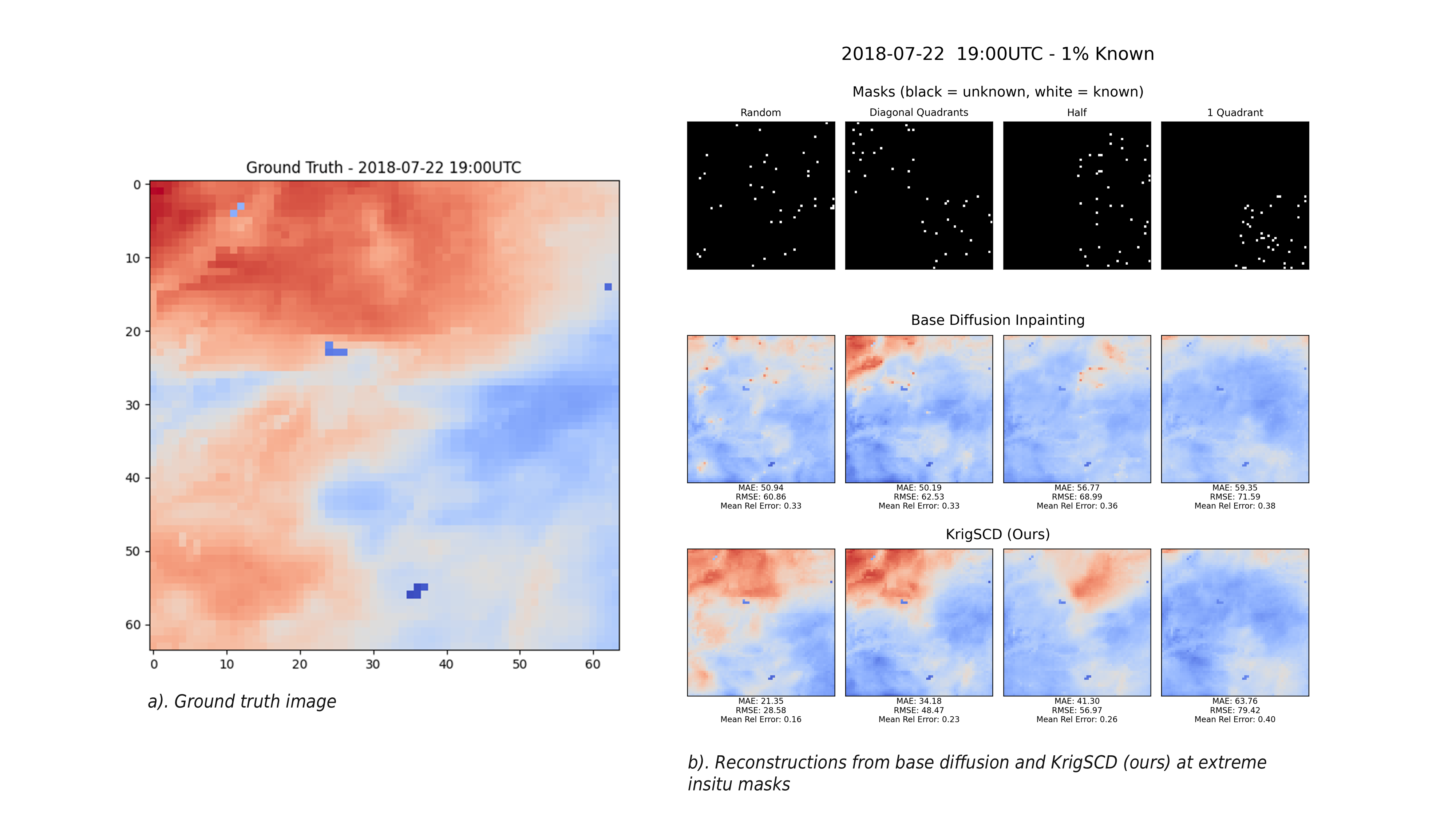}
    \caption{(a) Ground truth sample of 2018-07-22 19:00UTC on the right, (b) Visual demonstration of our algorithm's performance (3rd row) against base diffusion (2nd row) for corresponding binary masks representing known data in white and unknown data in black (1st row).}
\label{fig:pk_npk_results}
\end{figure}

We demonstrate the efficacy of our method on extreme masks of 1\% known, showing the drastic bump in improvement from base diffusion to KrigSCD. Figure \ref{fig:pk_npk_results} presents the ground truth for a given sample on the left, with reconstructed images from assorted masks of known insitu data on the right. To elaborate in greater detail, the masks in the first row denote the pixels of known and unknown observations that the model has access to when inpainting. All white pixels indicate a known observation, meaning that the model has access to the ground truth value at this location, whereas all black regions indicate "unknown," meaning that the model must spatially interpolate this area on its own. It's clear that the outputs from base diffusion in the second row properly take in the known pixel value (i.e. we do see agreement between the known pixel locations and the ground truth image on the left). However, other than those exact pixel locations, the base diffusion formulation struggles to capture the underlying dynamics of the ground truth. We hypothesize that this is primarily due to the UNet architecture of the diffusion model. Standard UNets use successive stride‑2 pooling, which acts like a low‑pass filter and exponentially attenuates the high‑frequency ``spike'' coming from the few isolated known pixels, so most conditioning is lost before the decoder sees it. As a result, pooling without anti‑aliasing also folds the mask’s broadband spectrum into the low‑frequency band, making it impossible to distinguish true signal from alias artifacts later in the network.

On the other hand, the one to one comparison with KrigSCD in the third row displays a much more spatially consistent reconstruction to the ground truth. The most evident case for this is shown with the first mask, where known observations are randomly scattered. Even with just 1\% of known data, KrigSCD manages to reconstruct the general structure of higher temperature regions—capturing the red areas at the top and beginning to suggest elevated values in the lower left corner. Base diffusion, on the other hand, fails to recover this structure, rendering the lower left corner almost entirely blue.

\begin{figure}[H]
    \centering
    \includegraphics[scale=0.6]{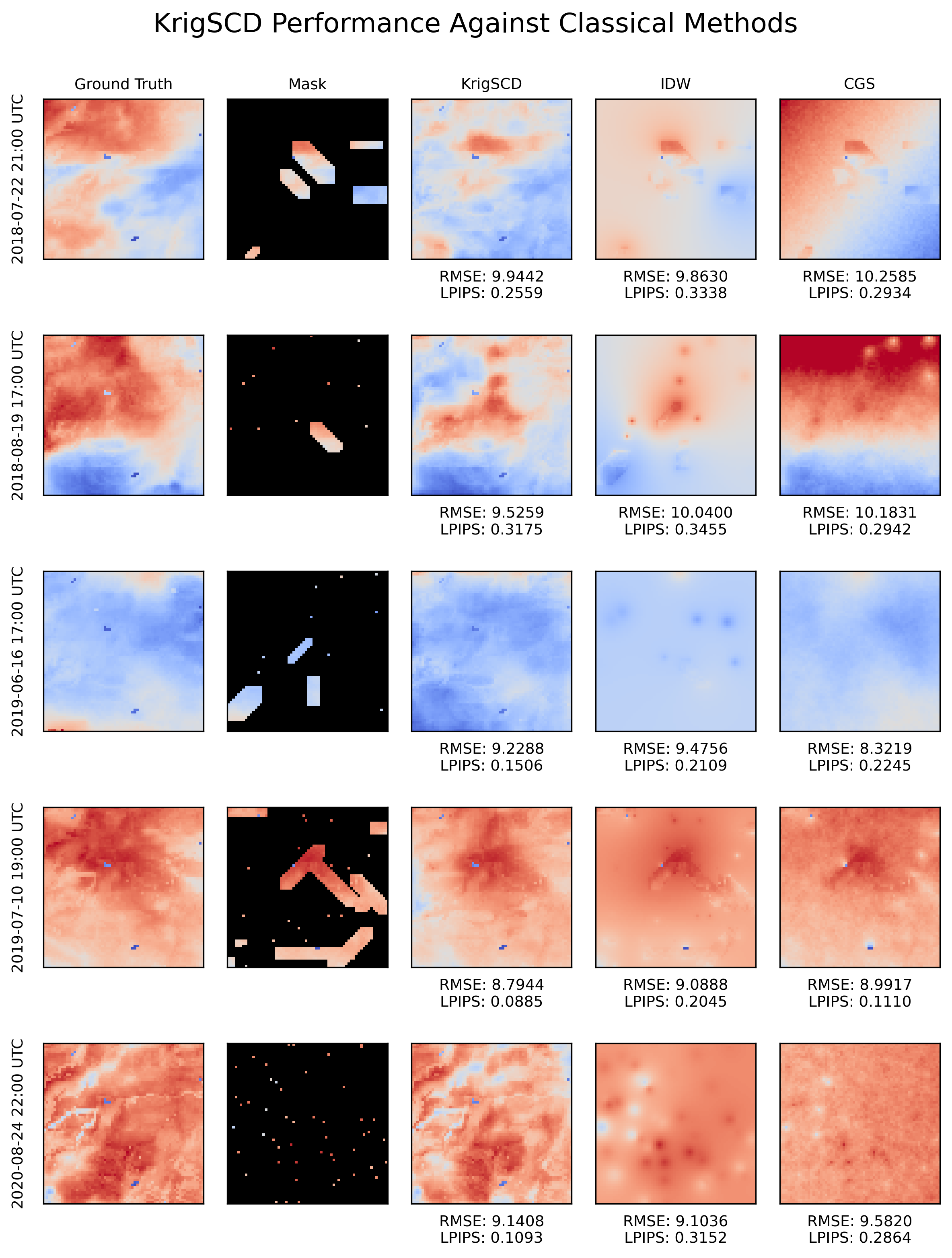}
    \caption{Reconstruction performance of the ensemble diffusion model output (3rd column) versus inverse distance weighting (4th column) and CGS (5th column) at five arbitrarily chosen test samples. The results indicate that for most cases the RMSE and LPIPS metric are minimized with ensemble KrigSCD. Visually examining the images and comparing them against the ground truth (1st column) reveal how well the model is capturing granular features of the data, even when only given incomplete coverage based on the corresponding mask (2nd column). This behavior is perhaps most evident in the 2020-08-24 sample shown in the last row.}
    \label{fig:composite}
\end{figure}

Moving forward, we visualize 5 arbitrarily chosen samples from our test set in Figure \ref{fig:composite} alongside traditional spatial interpolation methods. These samples are meant to represent a diverse spread of temperature fields and mask configurations to illustrate the performance of our framework in various situations. Here, the known observations within the mask are shown with their ground truth values to help aid the understanding of what is known vs. unknown. 

Based on pure visual inspection, the ability of KrigSCD to produce features consistent with the ground truth is impressive. Remarkably, this holds true even in cases where ensemble KrigSCD registers higher RMSE or LPIPS scores, such as in the third and second rows respectively, suggesting that purely quantitative metrics may not fully capture the perceptual fidelity of the reconstructions. Indeed, the human eye can still trace meaningful structures in KrigSCD's outputs that are largely absent in those of standard methods. For instance, stark, localized dark blue regions indicative of urban centers are clearly delineated by KrigSCD, yet remain undetected by both IDW and CGS. For further details regarding the algorithm and implementation of these traditional methods, please refer to the Appendix A.

We call the reader's attention to the last row of images. For this sample, the ground truth has an interesting dynamic that allows us to distinctly observe elaborate patterns within the data. Ensemble KrigSCD successfully reconstructs much of this intricate structure, while the other approaches fail to capture its complexity, offering only coarse or distorted approximations in comparison.

\begin{figure}[H]
    \centering
    \includegraphics[scale=0.6]{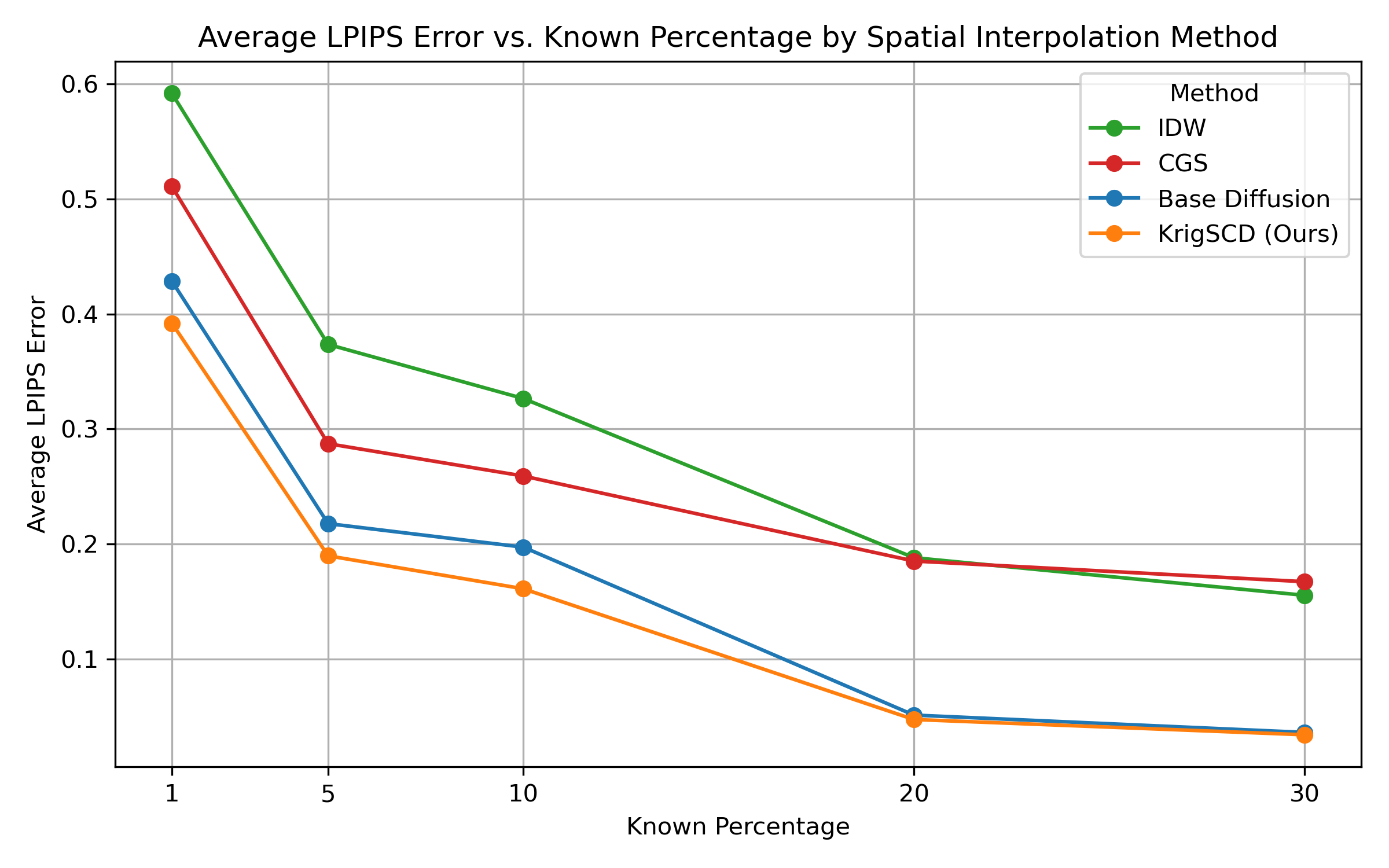}
    \caption{LPIPS scores for each method across varying known percentages. We note that the diffusion-based approaches produce lower errors in every case, and that after 20\% known we achieve LPIPS scores that indicate the reconstructions are virtually indistinguishable from the ground truth.}
    \label{fig:lpips}
\end{figure}

We assess perceptual reconstruction quality across interpolation methods using the LPIPS metric, which quantifies visual similarity in a manner aligned with human perception. Figure \ref{fig:lpips} presents LPIPS values for base diffusion, ensemble KrigSCD, inverse distance weighting (IDW), and CGS, evaluated over increasing proportions of known data. Each point reflects an average across 100 reconstruction samples per method and mask configuration. Notably, ensemble KrigSCD consistently yields lower LPIPS scores, indicating superior structural alignment with the ground truth. These results underscore KrigSCD’s ability to recover high-fidelity spatial patterns, even under sparse observation regimes. The magnitude of this improvement is formally captured in Table \ref{tab:krigscd_percent_comparison}, which reports the percentage gain in performance over baseline interpolation schemes, averaged across all evaluation metrics.

\begin{table}[h]
\centering
\begin{tabular}{cccc}
\topline
Mask \% Known & IDW & Kriging & Base Diffusion\\
\midline
\ \ 1&  33.77\% & 23.28\% & 8.52\%\\
\ \ 5& 49.21\% & 33.95\% & 12.86\% \\
\ \ 10& 50.68\% & 37.85\% & 18.35\%\\
\ \ 20& 74.82\% & 74.43\% & 7.48\%\\
\ \ 30& 78.07\%& 79.62\% & 5.66\%\\
\botline
\end{tabular}
\caption{LPIPS percentage improvement of KrigSCD over other traditional methods.}\label{tab:krigscd_percent_comparison}
\end{table}

\begin{figure}[H]
     \centering
     \noindent\includegraphics [width=30pc,angle=0]{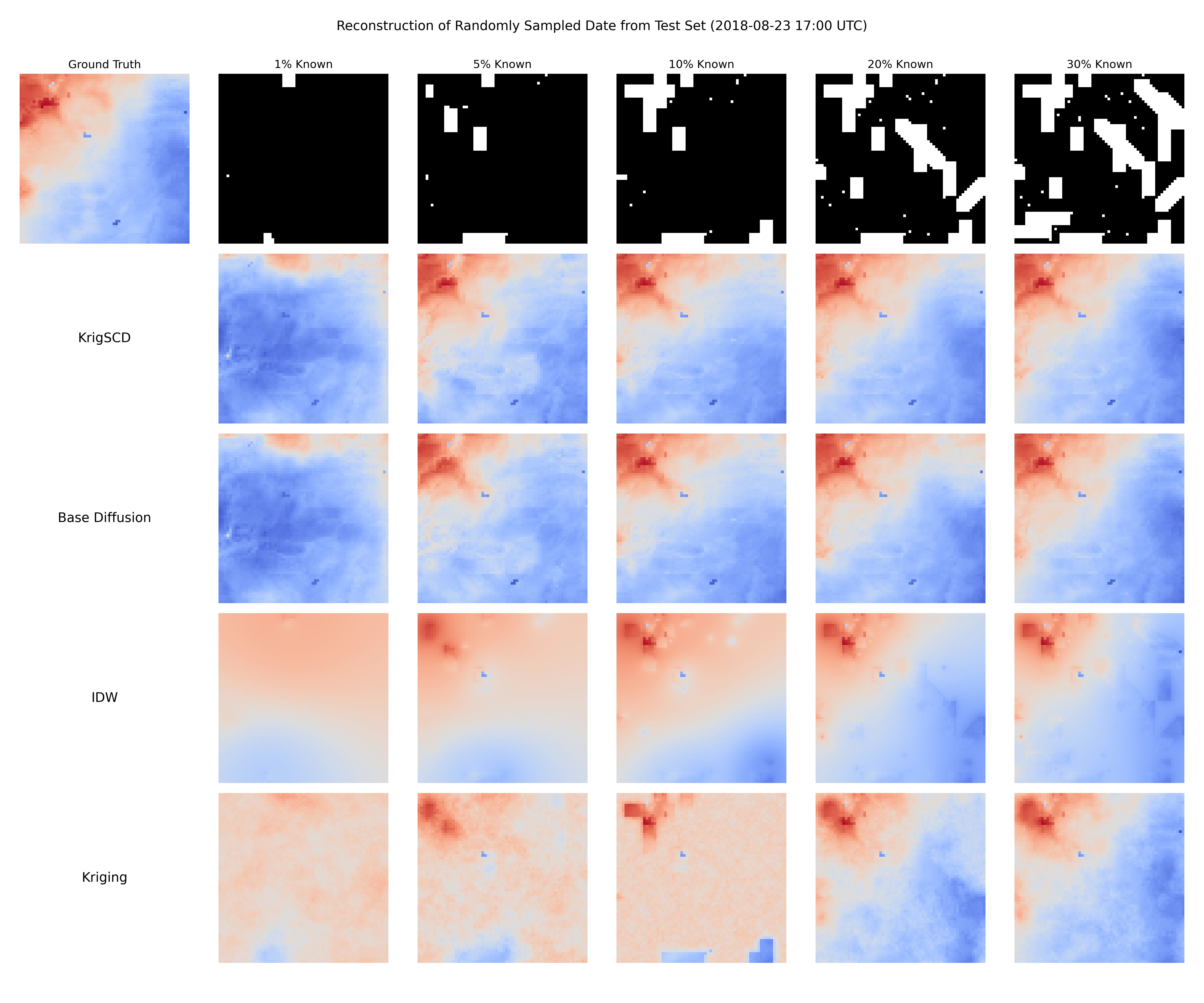}\\
     \caption{Reconstruction from each spatial interpolation method for a randomly selected sample (2018-08-23 17:00 UTC) from the test set. Each row after the first shows a realization from a different method.}\label{fig: distribution_masks}
\end{figure}

\begin{figure}[H]
     \centering
     \noindent\includegraphics [width=30pc,angle=0]{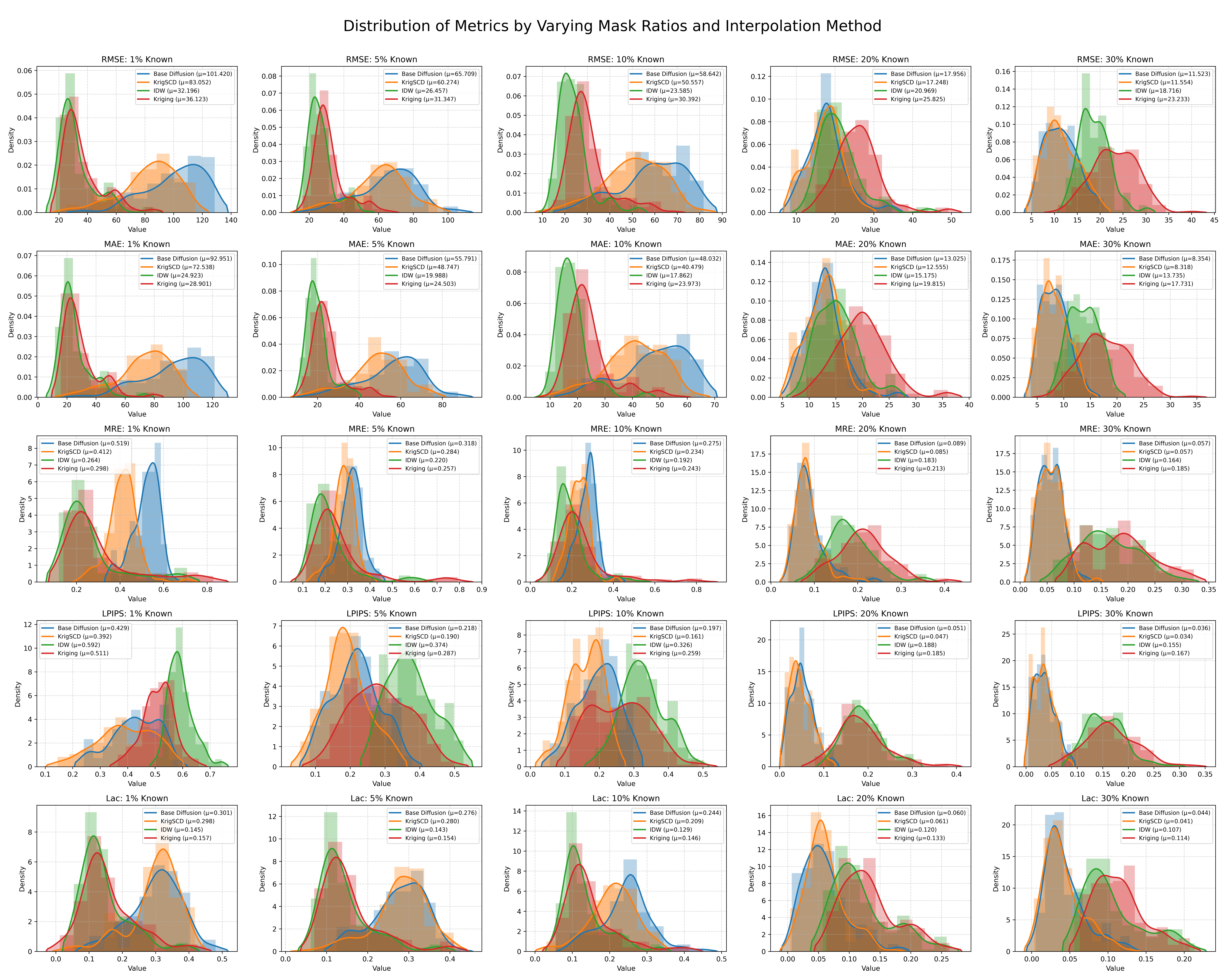}\\
     \caption{Metric distribution landscape across increasing levels of mask known percentages.}\label{fig: distribution_error}
\end{figure}

Figure \ref{fig: distribution_error} fully summarizes the distribution of metric performance across five different levels of mask coverage. It encapsulates the accumulation of our extensive results, with the KrigSCD algorithm run on 100 test images and 10 ensemble members for the base diffusion model, KrigSCD, and CGS. Note that these masks progressively build off of each other and are consistent across all 100 reconstructions. To illustrate a representative example from the test set, we randomly select one sample (2018-08-23 17:00 UTC) and visualize its reconstructions by method in Figure \ref{fig: distribution_masks}. This image is one of the 100 used in generating the overall distribution shown in Figure \ref{fig: distribution_error}, and its individual metric values constitute a single data point within that distribution.

As seen in Figure \ref{fig: distribution_error}, we note a few trends:
\begin{itemize}
    \item In every metric and every known percentage, KrigSCD outperforms the non-smoothed diffusion model.
    \item The diffusion-based methods outperform the classical methods slightly for lower percentage knowns (1-10\%) on the LPIPS metric but significantly overperform on higher percentages (20-30\%).
    \item RMSE/MAE values for classical methods are smaller for low percentage knowns, but this also makes sense due to the nature of their algorithms (intending to minimize the MSE). However, note that this does not necessarily produce images that look visually more plausible, as evidenced by the LPIPS score.
    \item The classical methods have a relatively constant lacunarity norm error across all known percentages. In contrast, the diffusion-based methods appear to converge to zero. While this metric isn't a perfect assessment for how well the methods capture the intrinsic gaps and patterns within the data, a decreasing error does indicate that the diffusion model is actively attempting to capture this information. This can be best seen in the fifth row of Figure \ref{fig:composite}.
\end{itemize}

Table \ref{tab: metrics} lists the independent average values for each metric by method, across the different percentages of known data. These numbers come directly from Figure \ref{fig: distribution_error}, but additionally we provide a mean of all independent values for any given method and metric, consolidating them in the rightmost column. This effectively tells a summary of the metric value that each method achieves for the percentages of known we investigate (1\%, 5\%, 10\%, 20\%, 30\%). While RMSE, MAE, and MRE seems to provide the lowest value for IDW (shown in red) and CGS (shown in green), we note how these metrics are made to minimize pixel-to-pixel error and do not capture overall broad scale spatial dynamics that are relevant for our application. On the other hand, LPIPS, a metric which captures this through the lens of human perception, is consistently lower for KrigSCD than for any other classical method.

\begin{table}[H]
    \centering
  \begin{tabular}{c|ccccc|c}
  \topline
    \multirow{2}{*}{\parbox{1.5cm}{\centering Metric}}
      & \multicolumn{5}{c|}{Percentage Known} 
      & \multirow{2}{*}{Average} \\ 

      & 1\% & 5\% & 10\% & 20\% & 30\% & \\ 
    \midline
    RMSE  &
      \begin{tabular}{@{}c@{}}\textcolor{blue}{101.420}\\\textcolor{orange}{83.052}\\
                          \textcolor{green}{32.196}\\\textcolor{red}{36.123}\end{tabular} &
      \begin{tabular}{@{}c@{}}\textcolor{blue}{65.709}\\\textcolor{orange}{60.274}\\
                          \textcolor{green}{26.457}\\\textcolor{red}{31.347}\end{tabular} &
      \begin{tabular}{@{}c@{}}\textcolor{blue}{58.642}\\\textcolor{orange}{50.557}\\
                          \textcolor{green}{23.585}\\\textcolor{red}{30.392}\end{tabular} &
      \begin{tabular}{@{}c@{}}\textcolor{blue}{17.956}\\\textcolor{orange}{17.248}\\
                          \textcolor{green}{20.969}\\\textcolor{red}{25.825}\end{tabular} &
      \begin{tabular}{@{}c@{}}\textcolor{blue}{11.523}\\\textcolor{orange}{11.5554}\\
                          \textcolor{green}{18.716}\\\textcolor{red}{23.233}\end{tabular} &
       \begin{tabular}{@{}c@{}}\textcolor{blue}{$51.075 \pm 2.222$}\\\textcolor{orange}{$44.582 \pm 1.852$}\\
                          \textcolor{green}{$24.388 \pm 0.558$}\\\textcolor{red}{$29.388 \pm 0.625$}\end{tabular} \\  
    \hline
    MAE &   \begin{tabular}{@{}c@{}}\textcolor{blue}{92.952}\\\textcolor{orange}{72.539}\\
                          \textcolor{green}{24.923}\\\textcolor{red}{28.901}\end{tabular} &
      \begin{tabular}{@{}c@{}}\textcolor{blue}{55.791}\\\textcolor{orange}{48.747}\\
                          \textcolor{green}{19.988}\\\textcolor{red}{24.503}\end{tabular} &
      \begin{tabular}{@{}c@{}}\textcolor{blue}{48.032}\\\textcolor{orange}{40.479}\\
                          \textcolor{green}{17.863}\\\textcolor{red}{23.973}\end{tabular} &
      \begin{tabular}{@{}c@{}}\textcolor{blue}{13.025}\\\textcolor{orange}{12.555}\\
                          \textcolor{green}{15.176}\\\textcolor{red}{19.815}\end{tabular} &
      \begin{tabular}{@{}c@{}}\textcolor{blue}{8.354}\\\textcolor{orange}{8.318}\\
                          \textcolor{green}{13.736}\\\textcolor{red}{17.731}\end{tabular} &
       \begin{tabular}{@{}c@{}}\textcolor{blue}{$43.654 \pm 2.073$}\\\textcolor{orange}{$36.568 \pm 1.634$}\\
                          \textcolor{green}{$18.340 \pm 0.458$}\\\textcolor{red}{$ 22.989 \pm 0.540$}\end{tabular} \\  
    \hline
    MRE  & \begin{tabular}{@{}c@{}}\textcolor{blue}{0.519}\\\textcolor{orange}{0.412}\\
                          \textcolor{green}{0.264}\\\textcolor{red}{0.298}\end{tabular} &
      \begin{tabular}{@{}c@{}}\textcolor{blue}{0.318}\\\textcolor{orange}{0.284}\\
                          \textcolor{green}{0.220}\\\textcolor{red}{0.257}\end{tabular} &
      \begin{tabular}{@{}c@{}}\textcolor{blue}{0.275}\\\textcolor{orange}{0.234}\\
                          \textcolor{green}{0.192}\\\textcolor{red}{0.243}\end{tabular} &
      \begin{tabular}{@{}c@{}}\textcolor{blue}{0.089}\\\textcolor{orange}{0.085}\\
                          \textcolor{green}{0.183}\\\textcolor{red}{0.213}\end{tabular} &
      \begin{tabular}{@{}c@{}}\textcolor{blue}{0.057}\\\textcolor{orange}{0.057}\\
                          \textcolor{green}{0.164}\\\textcolor{red}{0.185}\end{tabular} &
       \begin{tabular}{@{}c@{}}\textcolor{blue}{$0.252 \pm 0.0108$}\\\textcolor{orange}{$0.215 \pm 0.009$}\\
                          \textcolor{green}{$0.205 \pm 0.006$}\\\textcolor{red}{$0.239 \pm 0.007$}\end{tabular} \\  
    \hline
    LPIPS & \begin{tabular}{@{}c@{}}\textcolor{blue}{0.429}\\\textcolor{orange}{0.392}\\
                          \textcolor{green}{0.591}\\\textcolor{red}{0.511}\end{tabular} &
      \begin{tabular}{@{}c@{}}\textcolor{blue}{0.218}\\\textcolor{orange}{0.190}\\
                          \textcolor{green}{0.374}\\\textcolor{red}{0.287}\end{tabular} &
      \begin{tabular}{@{}c@{}}\textcolor{blue}{0.197}\\\textcolor{orange}{0.161}\\
                          \textcolor{green}{0.327}\\\textcolor{red}{0.259}\end{tabular} &
      \begin{tabular}{@{}c@{}}\textcolor{blue}{0.051}\\\textcolor{orange}{0.047}\\
                          \textcolor{green}{0.188}\\\textcolor{red}{0.185}\end{tabular} &
      \begin{tabular}{@{}c@{}}\textcolor{blue}{0.036}\\\textcolor{orange}{0.034}\\
                          \textcolor{green}{0.155}\\\textcolor{red}{0.167}\end{tabular} &
       \begin{tabular}{@{}c@{}}\textcolor{blue}{$0.186 \pm 0.010$}\\\textcolor{orange}{$0.165 \pm 0.009$}\\
                          \textcolor{green}{$0.327 \pm 0.010$}\\\textcolor{red}{$0.282 \pm 0.009$}\end{tabular} \\ 
    \hline
    Lac & \begin{tabular}{@{}c@{}}\textcolor{blue}{0.301}\\\textcolor{orange}{0.298}\\
                          \textcolor{green}{0.145}\\\textcolor{red}{0.157}\end{tabular} &
      \begin{tabular}{@{}c@{}}\textcolor{blue}{0.276}\\\textcolor{orange}{0.280}\\
                          \textcolor{green}{0.143}\\\textcolor{red}{0.154}\end{tabular} &
      \begin{tabular}{@{}c@{}}\textcolor{blue}{0.244}\\\textcolor{orange}{0.209}\\
                          \textcolor{green}{0.129}\\\textcolor{red}{0.147}\end{tabular} &
      \begin{tabular}{@{}c@{}}\textcolor{blue}{0.060}\\\textcolor{orange}{0.061}\\
                          \textcolor{green}{0.120}\\\textcolor{red}{0.133}\end{tabular} &
      \begin{tabular}{@{}c@{}}\textcolor{blue}{0.044}\\\textcolor{orange}{0.041}\\
                          \textcolor{green}{0.107}\\\textcolor{red}{0.114}\end{tabular} &
       \begin{tabular}{@{}c@{}}\textcolor{blue}{$0.185 \pm 0.008$}\\\textcolor{orange}{$0.178 \pm 0.008$}\\
                          \textcolor{green}{$0.129 \pm 0.004$}\\\textcolor{red}{$0.141 \pm 0.004$}\end{tabular} \\  
    \hline
  \end{tabular}
  \caption{Mean values for each metric.
    \textcolor{blue}{Blue}: base diffusion model;
    \textcolor{orange}{Orange}: KrigSCD;
    \textcolor{green}{Green}: IDW;
    \textcolor{red}{Red}: conditional gaussian simulations (CGS).}
  \label{tab: metrics}
\end{table}

\section{Parametric Study}
To evaluate the robustness and reliability of our framework against various changing parameters, we conducted a series of experiments varying key input parameters, including different spatial masking configurations, proportions of known data, and varying ratios of in situ to swath-based observations. These evaluations were designed to test the framework under a range of realistic and challenging conditions. The results from these experiments are presented to offer a comprehensive view of the strengths and limitations of KrigSCD, and to provide insight into its behavior across diverse spatiotemporal data scenarios. To that end, we present the results of our algorithm performance across two main changing factors: 
\begin{itemize}
  \item diverse splits between insitu vs. satellite swath observation types, and
  \item overall \% of known data at any given time.
\end{itemize}

\begin{figure}[H]
    \centering
     \noindent\includegraphics [width=30pc,angle=0]{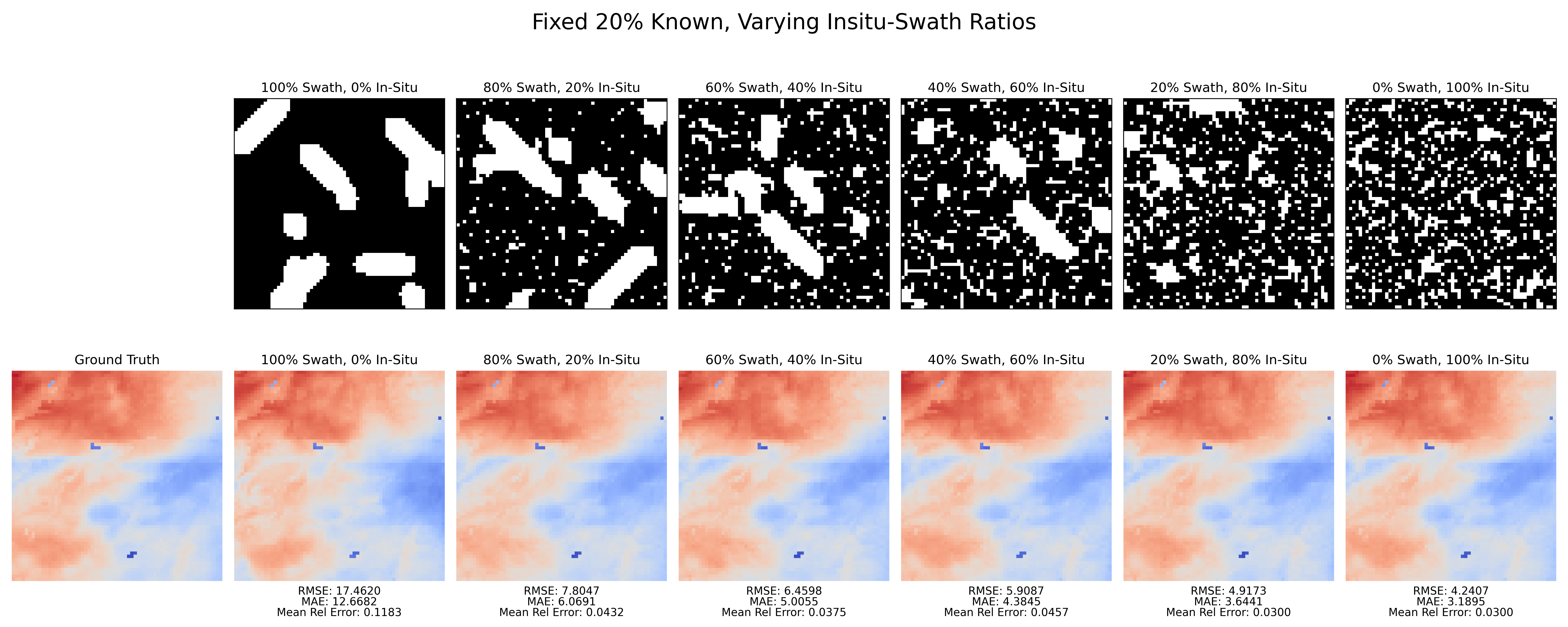}\\
     \caption{A comparison of the impact that varying insitu-swath ratios for a fixed 20\% known mask can have for a given test sample.}\label{fig: fixed20_both}
\end{figure}

Figure \ref{fig: fixed20_both} displays the model reconstructions across a fixed percentage of known data while varying the composition of observation modalities. From left to right, we note that as the ratio of swaths to insitu observations decreases, so does the RMSE, MAE, and MRE metric. These masks are specifically designed to be incremental in that they start from a given mask of all swaths and slowly adjusts to remove the same swaths while adding insitu observations until the specified ratio is reached. Doing this allows for greater interpretability across the study, so that the masks themselves remain as consistent across each run as possible. This design allows for a controlled analysis of performance impacts without introducing confounding variation in spatial coverage. The reason we observe monotonically decreasing values of RMSE can be attributed to the fact that KrigSCD capitalizes on randomly distributed, isolated data due to its kriging-smoothed algorithm. Even still, the overall structure of the temperature field is substantially captured with the 100\% swath-0\% insitu mask.

\begin{figure}[H]
    \centering
     \noindent\includegraphics [width=30pc,angle=0]{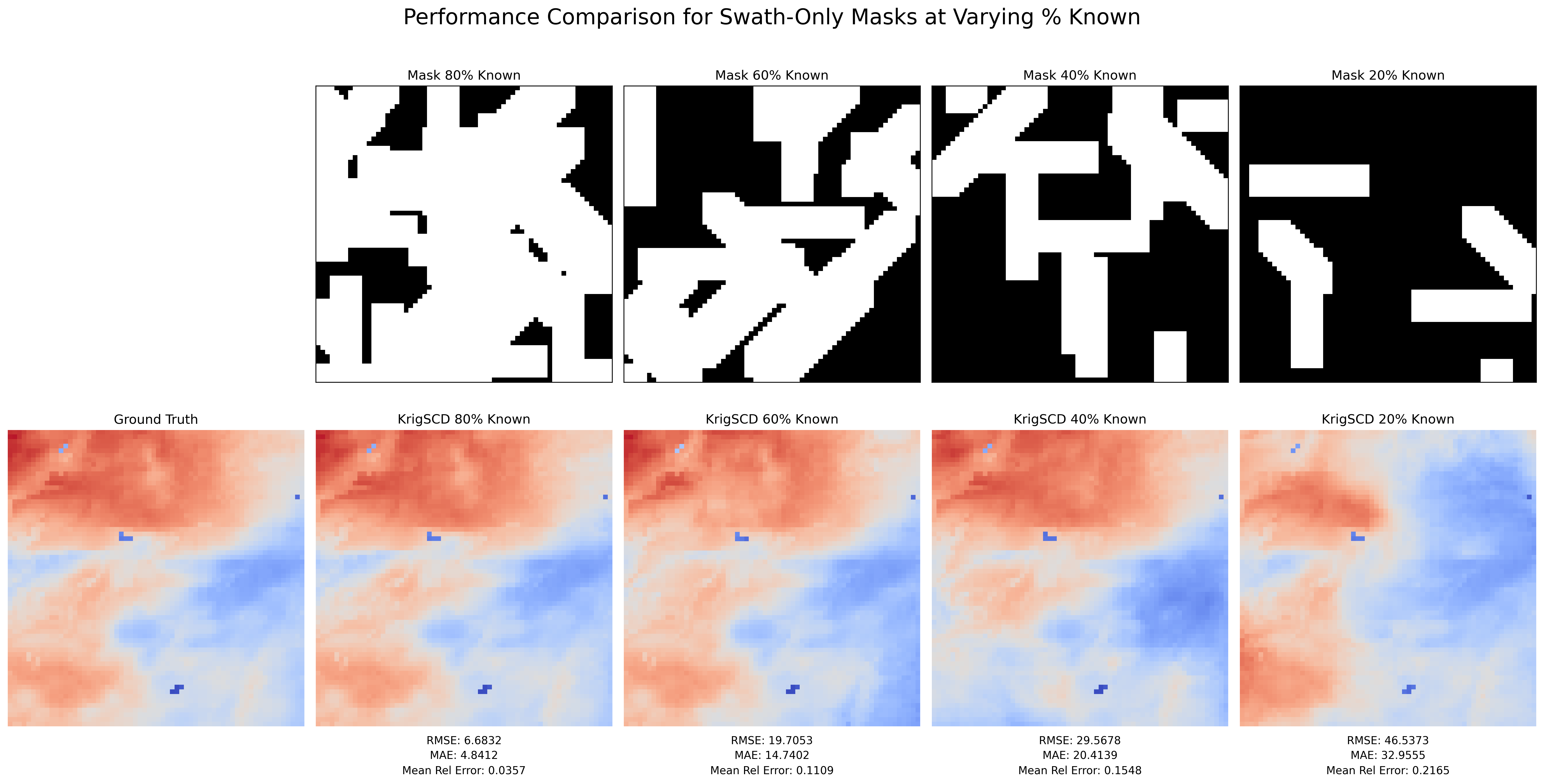}\\
     \caption{A study on masks with only swaths, ranging from 20\% to 80\% known percentages with 20\% increments. }\label{fig: swath_only}
\end{figure}

\begin{figure}[H]
     \centering
     \noindent\includegraphics [width=30pc,angle=0]{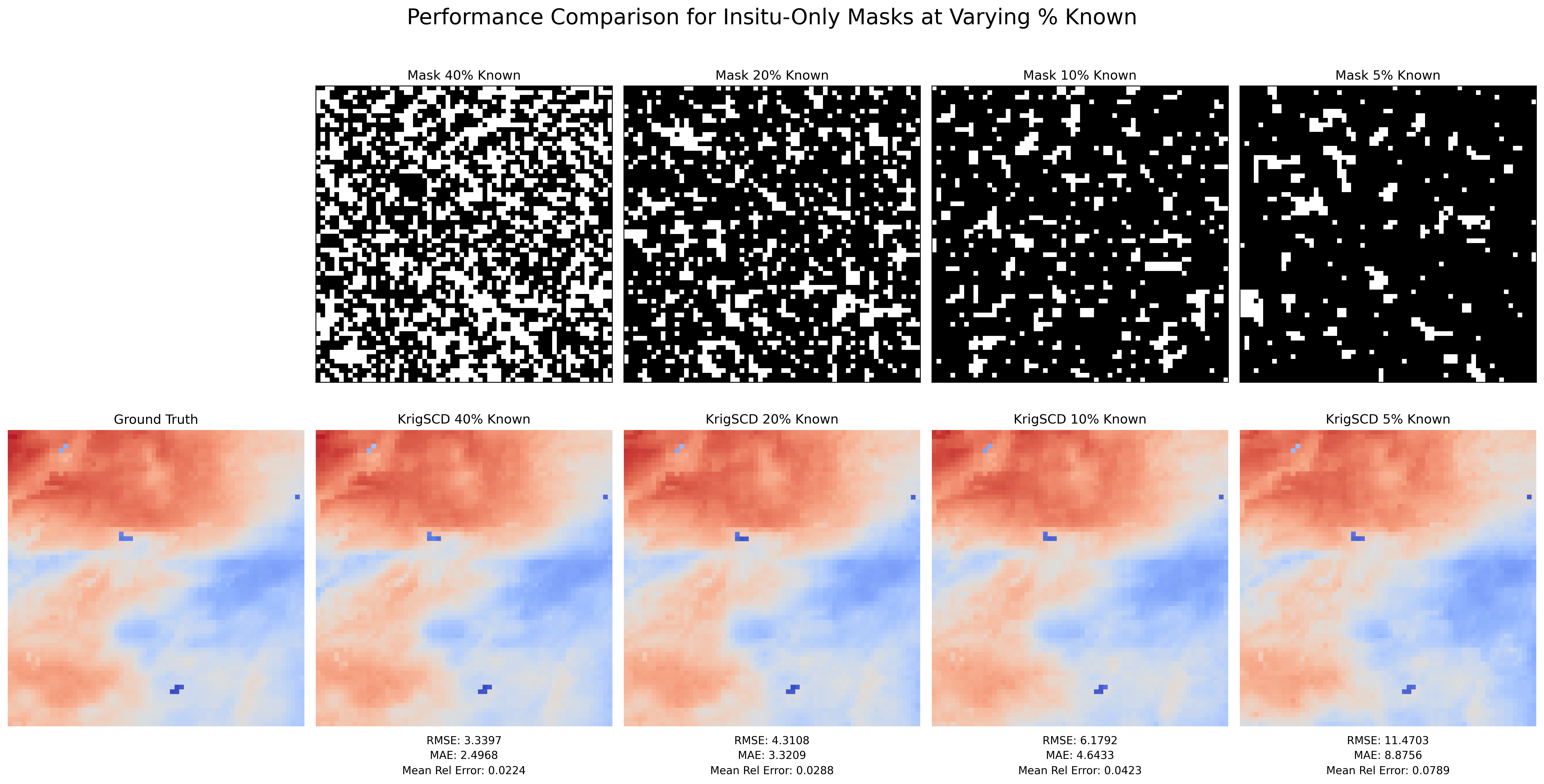}\\
     \caption{A study on masks with only in-situ observations, with known percentages at 5\%, 10\%, 20\%, and 40\%. }\label{fig: insitu_only}
\end{figure}

To further explore the boundaries of model performance and identify potential failure cases, we extend this analysis to exclusively swath-based and insitu-based masks at varying percentages of known. These are depicted respectively in Figures \ref{fig: swath_only} and \ref{fig: insitu_only}. In these investigations, we vary the overall coverage of known data from 20-80\% for the swath-only masks and 5-40\% for the insitu-only masks. This adjustment for finer granularity is intentional: applying the same 20–80\% range to insitu data would lead to trivially easy reconstructions due to the dense, randomized nature of the observations and the strong smoothing effect of the kriging process. As expected, increasing the percentage of known data consistently leads to lower RMSE, MAE, and MRE values, confirming the model’s capacity to scale its performance with data availability.

\section{Discussion \& Conclusion}
In summary, this work offers a high-level comparison between diffusion model-based approaches for conditional spatial interpolation and more traditional methods, while also highlighting the remarkable accuracy with which these models recover data attributes under conditions of extreme sparsity. We conducted an evaluation of the model's performance across various observation types and found it consistently capable of handling diverse measurement modalities with impressive flexibility.
More significantly, we provide provable evidence of the reconstruction capabilities of diffusion models when applied to arbitrary masking schemes using techniques from \citet{RePaint}, specifically those representing in-situ and swath measurements. This enables a far more continuous and coherent mapping of spatial data, especially in regions where conventional methods would either fail entirely or yield highly inaccurate estimates. Our approach incorporates a kriging-based smoother, which enhances the model’s ability to infer complex spatial characteristics which outperform IDW and standard kriging.

Although our results are demonstrated on temperature fields, we anticipate minimal barriers in extending this pipeline to other environmental variables such as wind, precipitation, and surface pressure. Perhaps most exciting is the potential application of this method as a robust initialization strategy for data assimilation (DA) models, and by extension, Earth System Models (ESMs). DA systems continue to grapple with the challenge of selecting appropriate starting conditions for reliable forecasting. We view our framework as a critical step toward incorporating deep generative models into the climate modeling pipeline, especially at the state estimation and prediction stages.

\subsection{Extension into Higher Dimensions}
\begin{figure}[H]
     \centering
     \noindent\includegraphics [width=30pc,angle=0]{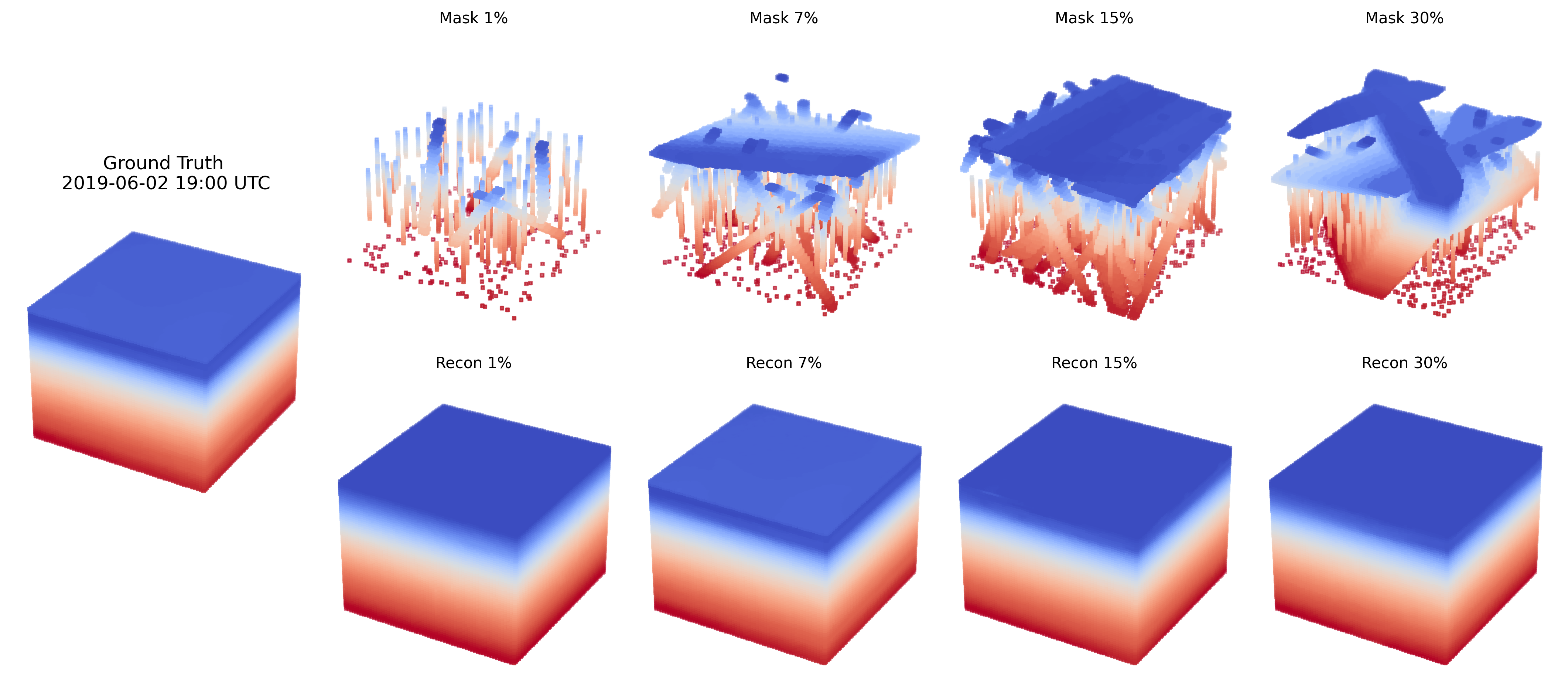}\\
     \caption{Visual performance of KrigSCD being implemented for 3D, where the top row shows the mask configurations (indicating the amount of known percentage we provide the model beforehand) and the bottom row shows the fully reconstructed sample.}\label{fig: 3d_results}
\end{figure}

As a natural extension to this work, we also present a way for KrigSCD to be applied for higher dimensionalities in Figure \ref{fig: 3d_results}. Here, we alter the design of the masks to incorporate modalities that are more consistent with a 3D space, including not just insitu and satellite swaths as in our 2D examples, but also weather balloons and plane swaths. The primary changes made to make this possible include performing convolutions and pooling operations in 3D, as well as building support for voxel data in the form of \texttt{.npy} files. The data was input as a 64x64x64 cube, and the number of head channels was adjusted accordingly. In addition, we employ the cosine variance scheduler within the diffusion model during the forward noising process instead of linear. As this is meant to simply demonstrate the potential for KrigSCD in increasing dimensions, we do not provide a thorough comparison with classical methods in the same dimensionality here.

\subsection{Limitations}
While impressive in functionality, the diffusion framework is not without its drawbacks. Most notably, these models require full-coverage input (i.e. no missing values) in order to properly train. Put simply, the basis for how these models work is still grounded in synthetic data, since HRRR and likely many other datasets are products of reanalysis, given that we will always have areas for which direct observations are not available. This leads to an interesting direction for which we hope to pursue in future works. Breaking the bounds of this full coverage bottleneck would mean fundamentally changing the way we interact with observations for the environment and have implications far beyond that of spatial interpolation, but also forecast and prediction.

Adjacent to this requirement, the model in its current state is space, time, and variable dependent. Namely, we would not expect it to perform well for other circumstances (e.g. say for inpainting missing wind data in California during winter months). Other roadblocks include the strict requirements for OpenAI's guided-diffusion codebase to work, some of which include (1) the requirement for observations to be projected onto pixels of discrete values between 0-255, which is problematic given that real-world climate data is continuous; (2) the computational power and time required to train and inpaint samples; and (3) as with most machine learning-related paradigms, a fair amount of effort is required to finetune hyperparameters to achieve sensible results.
\subsection{Future Work}
Looking forward, the avenues for extension are both varied and promising. One particularly compelling direction involves integrating physical constraints directly into the training objectives of neural networks. This approach, pioneered in recent work by \citet{10.1007/s10115-023-01829-2} and \citet{verma2024climode}, suggests that physics-informed loss functions could dramatically improve convergence rates and prediction accuracy. Given that one of the most cited disadvantages to DDPMs involve its required computational power, the idea of converting these paradigms to work in the context of latent diffusion models \citep{rombach2022highresolutionimagesynthesislatent} is also fascinating and relatively unexplored at the moment. This would enable the diffusion process to work in a lower dimensional latent space, learning the broader features of the image before projecting that back into pixel space for refinement with a VAE. Even more, the simultaneous learning of multiple variables with diverse sparsity levels and spatial resolutions presents a complex but high-impact challenge. Exploring this multidimensional problem could yield insights with far-reaching implications across climate science and beyond.

\clearpage
%%%%%%%%%%%%%%%%%%%%%%%%%%%%%%%%%%%%%%%%%%%%%%%%%%%%%%%%%%%%%%%%%%%%%
% ACKNOWLEDGMENTS
%%%%%%%%%%%%%%%%%%%%%%%%%%%%%%%%%%%%%%%%%%%%%%%%%%%%%%%%%%%%%%%%%%%%%
\acknowledgments
M.V. and V.T. acknowledge the support of the U.S. National Science foundation via research funding from National Science Foundation Projects CMMI-2042325 and 2332069.

%%%%%%%%%%%%%%%%%%%%%%%%%%%%%%%%%%%%%%%%%%%%%%%%%%%%%%%%%%%%%%%%%%%%%
% DATA AVAILABILITY STATEMENT
%%%%%%%%%%%%%%%%%%%%%%%%%%%%%%%%%%%%%%%%%%%%%%%%%%%%%%%%%%%%%%%%%%%%%
\datastatement
The HRRRv3 dataset used for this study is publicly available for download at the University of Utah's data archive: \url{https://home.chpc.utah.edu/~u0553130/Brian_Blaylock/cgi-bin/hrrr_download.cgi}. The corresponding code used to generate the main results can be found on GitHub: \url{https://github.com/valtsao17/krigSCD.git}.

%%%%%%%%%%%%%%%%%%%%%%%%%%%%%%%%%%%%%%%%%%%%%%%%%%%%%%%%%%%%%%%%%%%%%
% REFERENCES
%%%%%%%%%%%%%%%%%%%%%%%%%%%%%%%%%%%%%%%%%%%%%%%%%%%%%%%%%%%%%%%%%%%%%
\bibliographystyle{ametsocV6}
\bibliography{references}

%%%%%%%%%%%%%%%%%%%%%%%%%%%%%%%%%%%%%%%%%%%%%%%%%%%%%%%%%%%%%%%%%%%%%
% APPENDIXES
%%%%%%%%%%%%%%%%%%%%%%%%%%%%%%%%%%%%%%%%%%%%%%%%%%%%%%%%%%%%%%%%%%%%%

\appendix[A] \label{appendix: interp_methods}
\appendixtitle{Interpolation Methods}
\section{Ordinary Kriging}
In this section, we explain the process for the prekriging used in Section \ref{methodology}\ref{krigscd}.
\subsubsection{Empirical Semivariogram and Model Fitting} 
We first estimate the empirical semivariogram, which is defined as
\[
\gamma(h) = \frac{1}{2} \operatorname{Var}\left[z(\mathbf{x}) - z(\mathbf{x} + h)\right],
\]
where $h$ is the spatial lag. An exponential model is then used to fit the semivariogram:
\[
\gamma(h) = c \left( 1 - \exp\left(-\frac{h}{\tau}\right) \right),
\]
where $c$ is the sill (variance level) and $\tau$ is the range parameter that controls the rate of decay.

\subsubsection{Covariance Modeling}
Assuming stationarity, the covariance function between two points separated by a distance $h$ is given by
\[
C(h) = c \exp\left(-\frac{h}{\tau}\right).
\]
For an unknown pixel located at $\mathbf{x}_0$, define the covariance vector between $\mathbf{x}_0$ and the $n$ known pixels as
\[
C_i = c \exp\left(-\frac{\|\mathbf{x}_0 - \mathbf{x}_i\|}{\tau}\right), \quad i = 1, \dots, n.
\]
Similarly, the covariance matrix among the known pixels is
\[
\Sigma_{ij} = c \exp\left(-\frac{\|\mathbf{x}_i - \mathbf{x}_j\|}{\tau}\right), \quad i, j = 1, \dots, n.
\]

\subsubsection{Solving the System}
To interpolate the unknown pixel value, we solve the following augmented system to obtain the kriging weights $\mathbf{w}$ and the Lagrange multiplier $\lambda$, which enforces the unbiasedness constraint:
\[
\begin{bmatrix}
\boldsymbol{\Sigma} & \mathbf{1} \\
\mathbf{1}^T & 0
\end{bmatrix}
\begin{bmatrix}
\mathbf{w} \\
\lambda
\end{bmatrix}
=
\begin{bmatrix}
\mathbf{C} \\
1
\end{bmatrix},
\]
where $\mathbf{1}$ is an $n$-dimensional vector of ones. The estimated (kriged) value at $\mathbf{x}_0$ is then computed as a weighted sum of the known values:
\[
z^* = \sum_{i=1}^{n} w_i z_i.
\]
The kriging variance, which quantifies the interpolation uncertainty, is given by
\[
\sigma^2 = c - \mathbf{w}^T \mathbf{C} - \lambda.
\]
\section{Inverse Distance Weighting}
Another classical method that we compare our framework to is inverse distance weighting $-$ a deterministic method for interpolation where an unknown value at a target location is estimated as a weighted average of known values, and all weights are inversely related to the distance between the known and target point. This method assumes that points that are close to one another are more alike than those further apart. To formalize this, suppose we have a set of known data points $\{(x_i, y_i, z_i)\}_{i=1}^{N}$, where the coordinates $(x_i, y_i)$ represent the spatial locations and $z_i$ the corresponding data value at that point. For an unknown location $(x,y)$, the interpolated value $\hat{z}(x,y)$ is given by a weighted sum:
\[
\hat{z}(x,y) = \frac{\displaystyle \sum_{i=1}^{N} w_i(x,y) \, z_i}{\displaystyle \sum_{i=1}^{N} w_i(x,y)}.
\]
The weights \(w_i(x,y)\) are defined as:
\[
w_i(x,y) = \frac{1}{d_i(x,y)^p},
\]
where 
\[
d_i(x,y) = \sqrt{(x - x_i)^2 + (y - y_i)^2}
\]
is the Euclidean distance between the target point and the \(i\)-th known point, and \(p>0\) is the power parameter that controls the rate at which the influence of a known point decreases with distance. If the target location \((x,y)\) coincides exactly with one of the known locations, i.e., \(d_i(x,y)=0\) for some \(i\), then the interpolated value is taken directly as 
$\hat{z}(x,y) = z_i$, avoiding division by zero and preserving the known data exactly.
\section{Conditional Gaussian Simulations}
Let us define a spatial process $Z(\mathbf{s})$ at location $\mathbf{s}$. We then decompose it into a deterministic trend $m(\mathbf{s})$ (which we call the drift) and a stochastic component $\epsilon(\mathbf{s})$, which represents the spatially correlated residuals. The trend is found using ordinary least squares, while the residuals are modeled with an exponential semivariogram and simulated with conditional Gaussian simulation. 

Let $\{(\mathbf{s}_i, z_i)\}_{i=1}^N$ be the observed data at known locations, where $\mathbf{s}_i = (x_i, y_i) $ and $z_i$ is the measured value. Assume a linear model of the form $m(\mathbf{s})=a+bx+cy$. We find our coefficients by minimizing the SSE:
\[\min_{a,b,c} \sum_{i=1}^n [z_i-(a+bx_i+cy_i)]^2.\]
Then it is a well known result that 
\[\boldsymbol{\beta}=(\mathbf{A}^\intercal \mathbf{A})^{-1}\mathbf{A}^\intercal \mathbf{z},\]
where 
\[\mathbf{A}=\begin{bmatrix}
    1 & x_1 & y_1 \\
    1 & x_2 & y_2 \\ 
    \vdots & \vdots & \vdots \\
    1 & x_N & y_N
\end{bmatrix},\quad \boldsymbol{\beta}=\begin{bmatrix}
    a\\b\\c
\end{bmatrix},\quad \mathbf{z}=\begin{bmatrix}
    z_1\\z_2\\\vdots\\z_N
\end{bmatrix}.\]
Once we find $\boldsymbol{\beta}$, the residuals can be computed by taking $\mathbf{r}_i=z_i-m(\mathbf{s}_i)$, for $i=1,2,\dots, N$.

As for our residuals $\epsilon(\mathbf{s})$, we assume them to be a stationary Gaussian random field with zero mean, and their spatial structure is characterized by the semivariogram $\gamma(h)$, which is defined as
\[\gamma(h) = \sigma^2\left[1-\exp\left(-\frac{h}{\tau}\right)\right],\]
where $h$ is the separation distance, $\sigma^2$ is the variance of the residuals, and $\tau$ is a range parameter controlling the rate of spatial correlation decay. In practice, we compute the experimental semivariogram using all pairs of residuals:
\[\gamma_{ij}=\frac{1}{2}(r_i-r_j)^2.\]
These values are binned and averaged to provide an empirical semivariogram, which is then fitted to the exponential model using nonlinear least-squares optimization. Now to perform kriging of the residual field at an unsampled location $\mathbf{s}_0$, we assume that the covariance function is given by 
\[C(h) = \sigma^2 \exp\left(-\frac{h}{\tau}\right).\]
The simple kriging estimator at $\mathbf{s}_0$ is defined as a weighted sum of the residuals at the $n$ nearest neighboring observation locations:
\[\hat{\epsilon}(\mathbf{s}_0)=\sum_{i=1}^n \lambda_i r_i.\]
The weights $\boldsymbol{\lambda}=[\lambda_1,\lambda_2,\dots, \lambda_n]^\intercal$ are obtained by solving $\boldsymbol{\Sigma}\boldsymbol{\lambda}=\mathbf{c}$, where $\boldsymbol{\Sigma}$ is an $n\times n$ covariance matrix among the neighboring points with elements
\[\Sigma_{ij} = \sigma^2 \exp\left(-\frac{\|\mathbf{s}_i-\mathbf{s}_j\|}{\tau}\right),\]
and $\mathbf{c}$ is the covariance vector between the prediction location and the $n$ neighbors:
\[c_i = \sigma^2 \exp\left(-\frac{\|\mathbf{s}_0-\mathbf{s}_i\|}{\tau}\right).\]
The kriging variance is given as $\operatorname{MSE}(\mathbf{s}_0)=\sigma^2-\mathbf{c}^\intercal \boldsymbol{\lambda}$. Rather than using only the kriging predictor, we perform sequential Gaussian simulation to generate multiple realizations of the residual field that honor both the local statistics and spatial continuity. In the simulation procedure, the residuals are first standardized by letting
\[r_i^*=\frac{r_i-\mu_r}{\sigma_r},\]
with sample mean $\mu_r$ and standard deviation $\sigma_r$. A grid is defined over the spatial domain and for each grid cell that is not conditioned by an observation, the following steps are executed in a random sequential order:
\begin{enumerate}
    \item[(1)] For a grid point $\mathbf{s}_0$, identify the $n$ neighboring conditioned points within a maximum distance.
    \item[(2)] Apply the simple kriging system to obtain a conditional mean $\hat{r}^*(\mathbf{s}_0)$ and kriging variance $\operatorname{MSE}(\mathbf{s}_0)$.
    \item[(3)] Draw a simulated value from the Gaussian distribution 
    \[r^*(\mathbf{s}_0)\sim \mathcal{N}(\hat{r}^*(\mathbf{s}_0), \operatorname{MSE}(\mathbf{s}_0)).\]
    \item[(4)] Update the set of conditioned points by adding the simulated value at $\mathbf{s}_0$.
\end{enumerate}
After simulating on the standard normal scale, the field is transformed back to the original scale via
\[r(\mathbf{s})=r^*(\mathbf{s})\sigma_r+\mu_r.\]
The final regression kriging estimate is then obtained by summing the deterministic trend and the simulated residual field:
\[\hat{Z}(s)=m(\mathbf{s})+r(\mathbf{s}).\] 
This combined approach incorporates both the large-scale variation (through the regression trend) and the local spatial variability (through the conditional simulation of residuals).

\end{document}